\documentclass[10pt]{iopart}
\usepackage{amsmath}
\usepackage{amssymb}
\usepackage{amstext}
\usepackage[squaren,thickqspace,thickspace,cdot]{SIunits}
\usepackage{graphics}
\usepackage{graphicx}
\usepackage{psfrag}
\usepackage{setspace}

\newcommand{\epsinfa}{\epsilon}
\newcommand{\eps}{\varepsilon}
\newcommand{\dd}{{\textrm{d}}}
\newcommand{\longchain}{\protect\mbox{--- $\cdot$ ---}}

\begin{document}
\article[Dielectric mixtures]{submitted to IEEE Trans. on El. Insul. and Dielec.}{Dielectric mixtures -- electrical properties and modeling}
\author{En{\.{\i}}s Tuncer\dag, Yuriy V. Serdyuk{\ddag} and Stanis{\l}aw M. Guba{\'n}ski}
\address{Chalmers University of Technology, 412\ 96 Gothenburg Sweden}
\address{\dag {\tt enis.tuncer@elkraft.chalmers.se}}
\address{\ddag {\tt yuriy.serdyuk@elkraft.chalmers.se}}
\date{\today}
\begin{abstract}
A review of current state of understanding of dielectric mixture properties and approaches to use numerical calculations for their modeling  are presented.  It is shown that interfacial polarization can yield different non-Debye dielectric responses depending on the properties of the constituents, their concentrations and geometrical arrangements. Future challenges on the subject are also discussed.

\end{abstract}

\section{Introduction}

Many materials used in our daily life are composites, which are often made up of at least two constituents or phases. The outstanding mechanical properties of many composites, and especially the unique combination of low density with high strength and stiffness have led not only to extensive research but also to a highly developed technologies~\cite{Hull,Combinatorial,Gilormini}. In comparison, relatively little attention was given to their other physical properties, which in parallel affected their use in electrical applications~\cite{Hale1976,Landauer1978,Lux1993,SihvolaBook}. The main advantage of composites is the ability to tailor materials for special purposes. Their applications are evolving day by day through the developments which lead to better precision in their manufacturing.

Designing of composite materials for electrical applications with classical {\em trial and error} approach   requires a lot of time and money.  However, by using computers, modeling desired properties of insulation systems can be estimated and corrected. The electrical properties of the system, {\em i.e.}, its conductivity and dielectric permittivity, are influenced by the  properties of the constituents, interaction between them and geometrical configuration.  One should not forget that insulating materials or dielectrics, as Faraday called them~\cite{Faraday}, show various properties at different voltages, temperatures, frequencies, moisture content and mechanical stresses. These should be considered in the design as well as in the diagnostics. 

From the early days of the electromagnetic field theory, predicting and calculating the dielectric properties of mixtures has been a challenging problem of both theoretical and practical importance~\cite{SihvolaBook,Landauer1978,Hale1976,bat74,Lux1993,Feldman1,McPhedran1,Dias1,McLachlanChen}. In composites the polarization of charges due to the differences between the electrical properties of the facing constituents plays an important role. Maxwell~\cite{Maxwellbook} was the first to notice this phenomenon when he considered a binary layered structure and expressed the effective dielectric properties of the composite. When structures other than the layered arrangement of phases were studied~\cite{Wagner1913,Wagner1914,Landauer1978,Hale1976,Lux1993,SihvolaBook,Pier6}, it was realised that the problem was not trivial, and even the geometrical shapes~\cite{SihvolaBook,Sillars1937,Steeman1992,TuncerPhD,Helsing1} and arrangements of inclusions~\cite{Tuncer1998b,sar97,Helsing1,Carrique,Luciano,men84,Lakhtakia2,YuYuen} played an important role in the electrical properties of composites. 

The aim of this article is to give a brief review on dielectric mixture systems studied in the literature and to attract attention to the behavior of interfacial polarization in frequency dependent fields. First, dielectrics and polarization phenomenon are presented in the next section. Later sections summarize different theoretical studies avaliable in the literature as well as numerical simulations performed by the authors on {\em ideal} structures. The simulations presented were in two dimensions and were designed such that the chosen material parameters have yielded dielectric relaxations at frequencies lower than $1\ \mega\hertz$.  Simulations of three dimensional structures are so far very rear~\cite{Brosseau1,Boudida2,Brosseau2,Boudida,Sareni1} and there is still a need for developments.

\section{Dielectrics}
\label{sec:Maxwell}

\subsection{Polarization in dielectrics}
The interaction between electromagnetics fields and matter is described by Maxwell equations. Vectors expressing the electrical components, dielectric displacement $\mathbf{D}$ and electric field $\mathbf{E}$ of the electromagnetic phenomena in dielectrics are interrelated.
\begin{gather}\begin{split}
\mathbf{D} = & \varepsilon_0 \mathbf{E} + \mathbf{P}%
\label{eq:dielectric_displacement}
\end{split}\end{gather}
The constant $\varepsilon_0$ is the permittivity of the free space, $1/36\pi\ \nano\farad/\meter$. The polarization vector, $\mathbf{P}$,
can also be written as follows,
\begin{gather}\begin{split}
\nabla \cdot \mathbf{P} = & -\rho\label{Poldiff}
\end{split}\end{gather}
where $\rho$ is the charge density.
The relation between the dielectric displacement, $\mathbf{D}$, and the applied field, $\mathbf{E}$, is often linear and can be expressed with a simple proportionality constant, $\varepsilon$,
\begin{gather}\begin{split}
\mathbf{D} = & \varepsilon \varepsilon_0 \mathbf{E}
\label{eq:D=epsE}
\end{split}\end{gather}
The constant $\varepsilon$ is called the relative permittivity, which describes the dielectric properties of medium.
When the polarization, $\mathbf{P}$, is taken into account (inserting Eq.~(\ref{eq:D=epsE}) in Eq. (\ref{eq:dielectric_displacement})), it is also proportional to the field, $\mathbf{E}$
\begin{gather}\begin{split}
\mathbf{P} = & \chi \varepsilon_0\mathbf{E} \equiv \varepsilon_0( {\varepsilon-1}) \mathbf{E}\label{eq:PvsE}
\end{split}\end{gather}
The quantity $\chi$ is called the polarization coefficient of the substance, or its dielectric susceptibility. Observe that in free space, $\varepsilon=1$, there is no polarization and $\chi=0$.

\begin{figure}[t]
   \centering{\includegraphics[height=3.in,width=2in,angle=-90]{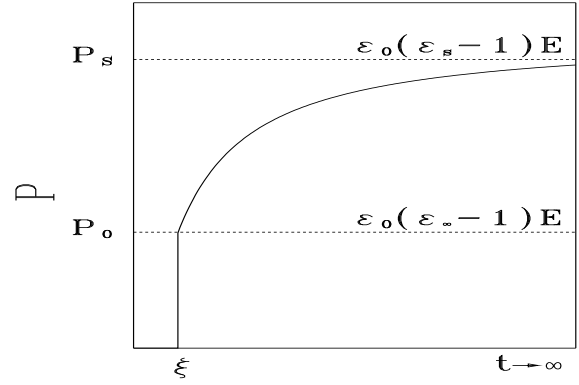}}
\caption{\label{fig:polarization.vs.time}
Time dependence of polarization, $P$, when a constant electric field is applied at $t=\xi$.}
\end{figure}

The polarization in materials can be due to several mechanisms; electronic, ionic (molecular), atomic, dipolar (orientational), and interfacial polarizations~\cite{Jonscher1983,Scaife,Frohlich,Bottcher}. Furthermore, hopping of charge carries between localized sites also creates polarization~\cite{Jonscher1983,jons92,Dyre,Dyre_a,Frohlich}. The first three polarization mechanisms are much quicker than the others. For this reason, depending on the time scale, the fast polarizations can together be considered as an instantaneous polarization, $\mathbf{P_0}$. Moreover, the polarization is to be finite at longer times, $\mathbf{P}(t\rightarrow\infty)=\mathbf{P}_{\textrm{s}}$, since infinite amount of electrical energy can not be stored in a dielectric (or a capacitor). Fig.~\ref{fig:polarization.vs.time} shows the polarization for a medium when a constant-step electric field, $\mathbf{E}=E_0\mathbf{\hat{n}}$ with  $\mathbf{\hat{n}}$ being the unit vector, is applied at $t=\xi$. The permittivity values at $t\rightarrow \infty$ and $t-\xi\approx0$ are defined as $\varepsilon_s$ and $\varepsilon_\infty$ which are the static and instantaneous (high frequency) dielectric constants, respectively.   
The time dependence of polarization is expressed as following when the electric history of the material is known
\begin{gather}
  \begin{split}
    \mathbf{P}(t)&=\varepsilon_0 (\varepsilon_\infty-1)\ \mathbf{E}(t) + \varepsilon_0 \int_{-\infty}^{t}{\sf f}(t-\xi)\ \mathbf{E}(\xi) d\xi
  \end{split}\label{eq:P.general.final}
\end{gather}
where ${\sf f}(t)$ is called the dielectric response function and it is an intrinsic material property. For harmonic fields Eq.~(\ref{eq:P.general.final}) and Eq.~(\ref{eq:dielectric_displacement}) yield to the complex dielectric permittivity, $\widetilde{\varepsilon}$, as follows.
\begin{gather}
  \begin{split}
    \widetilde{\varepsilon}(\omega)=& \varepsilon_\infty + \int_{0}^{\infty} {\sf f}(t) \exp(-\imath \omega t) \dd t
  \end{split}
\end{gather}
Therefore, the Fourier transform of the response function ${\sf f}(t)$ is defined as the complex dielectric susceptibility, 
\begin{gather}
  \begin{split}
\chi(\omega)&\equiv\widetilde{\varepsilon}-\varepsilon_{\infty}\\
    \chi(\omega)&=\chi'-\imath\chi''= \int_{0}^{\infty} {\sf f}(t) \exp(-\imath \omega t) \dd t
  \end{split}
\end{gather}
The real and imaginary parts of the dielectric susceptibility are coupled to each other by Kramer-Kronig relations \cite{LL}.

\subsection{Interfacial polarization}\label{sec:interf-polar}

When two media are put into contact (forming an interface) and an electric field is applied, charge polarization occurs at the interface due to the differences between the ratios of the electrical properties (conductivity and permittivity). This phenomena was first studied by Maxwell~\cite{Maxwell1954} who considered two phase system with one of the phases insulating, $\sigma_1=0$. The system was a layered strucure and the effective dielectric permittivity of the mixture was expressed as a function of intrinsic electrical properties of the phases, $\sigma_{1,2}$ and $\epsilon_{1,2}\equiv\varepsilon_{\infty_{1,2}}$ and their volume fraction $q$. The complex dielectric permittivity of the phases, $\widetilde{\varepsilon}$, in that case is
\begin{gather}\begin{split}
  \label{eq:compl_eps}
    \widetilde\varepsilon_1(\omega)&=\epsilon_1\\
    \widetilde\varepsilon_2(\omega)&=\epsilon_2+\frac{\sigma_2}{\imath\varepsilon_0\omega},
\end{split}\end{gather}
and the effective complex permittivity of the mixture can be expressed as 
\begin{equation}
  \label{eq:maxw1}
  \widetilde\varepsilon_{\sf e}(\omega)=\epsilon_{\sf e} + \frac{\Delta\varepsilon}{1+\imath \omega \tau} 
\end{equation}
where $\Delta\varepsilon$ is the dielectric strength of the interfacial polarization and $\tau$ is the relaxation time of the polarization. This expression is in the form of Debye relaxation function~\cite{Debye1945}.
\begin{gather}\begin{split}
 \label{eq:maxwDeb}
 \Delta\varepsilon=\frac{q\epsilon_{1} \epsilon_{\sf e}}{(1+q)\epsilon_{2}}\quad \text{and} \quad
 \tau=\frac{q(\epsilon_{1} + \epsilon_{2})+\epsilon_{2}}{(1+q)\sigma_{2}},
\end{split}\end{gather}
and $ \epsilon_{\sf e} $ is the effective dielectric constant at optical (high) frequencies
\begin{gather}\begin{split}
  \label{eq:Maxw}
 \epsilon_{\sf e}=\frac{\epsilon_{1} \epsilon_{2}(1+2q)}{q(\epsilon_{1}+ \epsilon_{2})+\epsilon_{2}}
\end{split}\end{gather}
A more general case is that both media have complex dielectric permittivities,  $\widetilde\varepsilon_1$ and $\widetilde\varepsilon_2$, in this case the effective dielectric permittivity is expressed as~\cite{Maxwellbook,TuncerPhD}
\begin{equation}
  \label{eq:maxw2}
   \widetilde\varepsilon_{\sf e}(\omega)=\frac{ \widetilde\varepsilon_1  \widetilde\varepsilon_2}{q \widetilde\varepsilon_1 + (1-q) \widetilde\varepsilon_2}
\end{equation}
The polarized charge density in the interface $\rho$ can be calculated from the Maxwell equations, and is expressed as 
\begin{equation}
  \label{eq:pol_charge}
  \rho(\omega)\propto\left[\frac{\epsilon_1 \widetilde\varepsilon_2 - \epsilon_2 \widetilde\varepsilon_1}{q\widetilde\varepsilon_1+(1-q)\widetilde\varepsilon_2}\right]
\end{equation}
The approach of Maxwell (two layer structure) is trivial and yields the only fully valid analytical formulas avaliable in the dielectric mixture theory for any composition of constituents. It can easily be extented to multi-layer systems.

\subsection{Electrical conduction}\label{sec:Electr-cond}

In reality, there are no perfect dielectrics, where conductivity is not present. The resulting total current density flowing through a dielectric when a step electric field is applied, can be written as,
\begin{gather}
  \begin{split}
    \mathbf{j}(t)=&\frac{\partial\mathbf{D}(t)}{\partial t}+\sigma\mathbf{E}(t)\\
              =&\varepsilon_0 [\varepsilon_\infty\delta(t)+{\sf f}(t)]\mathbf{E}(t)+\sigma\mathbf{E}(t)\\
  \end{split}\label{eq:current.j}
\end{gather}
The two asymptotic parts of the current density, $\mathbf{j}$, are the instantaneous current density due to the capacitive component, $\varepsilon_0\varepsilon_\infty\delta(t)$, and the dc conduction current density due to the conductivity, $\sigma$, of the material, respectively. The current density due to the materials polarization is given by the response function, ${\sf f}(t)$.

The conduction in materials may be real dc conduction which can be divided into two classes: band conduction \cite{AshcroftandMermin,Sutton} and  dc  hopping conduction \cite{Elliot_Amorphous,Zallen}. The band conduction is present in the absence of defects and impurities, and it is led by the band structure of the material. 
The latter, dc hopping conduction, takes place via defects or impurities which form potential wells (traps or localized states) that are favorable for charge carriers (electrons, holes and ions) to hop. It is therefore a phenomenon in disordered materials. 

\begin{figure}[tbp]
  \begin{center}
\psfragscanon
\psfrag{R}[c][]{$R$}
\psfrag{E1}[c][]{$E_1$}
\psfrag{E2}[c][]{$E_2$}
\psfrag{0}[c][]{0}
\psfrag{E}[c][]{$E$}
   \includegraphics[width=3.3in]{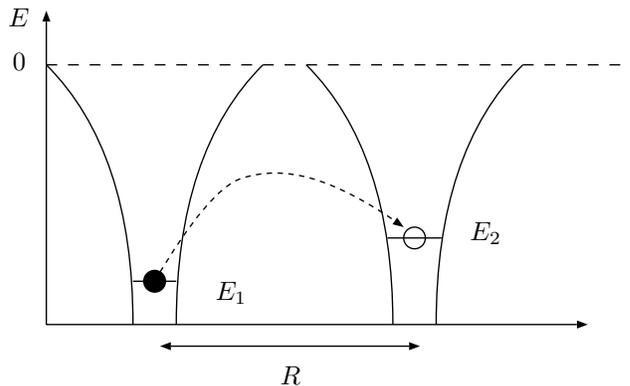}
\psfragscanoff
    \caption{Charge transport between hopping sites.}
    \label{fig:hopping2sites}
  \end{center}
\end{figure}
The first approach to hopping conductivity was described by Mott and Davis~\cite{Mott}. They had improved the work of Anderson~\cite{Anderson1958,Plischke1994} on localization (trapping) of electrons. The hopping conduction model of Mott and Davis showed that, at a finite temperature, electrons are able to tunnel between localized states as in Fig.~\ref{fig:hopping2sites} (for further reference see \cite{Elliot_Amorphous}). The energy and position distributions of the hopping sites decide the transition rate, $p_{1-2}$~\cite{Mott,Zallen},
\begin{equation}
  \label{eq:hoppingrate}
  p_{1-2}\sim \exp\biggl [ -2\alpha R-\frac{\Delta}{k_\mathrm{B}T}\biggr ]
\end{equation}
where $\Delta$ is a function of the energies of the states $E_{1,2}$, and $k_\mathrm{B}T$ is the thermal energy; $k_\mathrm{B}$ is the Boltzmann constant, $k_\mathrm{B}=86.174\ \micro e\volt\per\kelvin$. Note that the charge carriers can in principle tunnel to any state depending on the product of inverse localization length $\alpha$ and hopping distance $R$. Although, the distance $R$ in Fig. \ref{fig:hopping2sites} is between two neighboring sites, the model uses the mean hopping distance. For a disordered system in zero electric field condition the rate of change of charges moving from site 1 to site 2 is  equal to the rate of change in charges moving from site 2 to site 1. When an electric field is applied, the rate of change will be biased to one side depending on the polarity of the applied field.

The localized states are disordered in insulators and amorphous semiconductors, when their spatial and energy  distributions are taken into consideration. Therefore, perturbations in energy of the localized states activate tunneling between the sites. Perturbations can be in the form of temperature, pressure or applied electric field. If the applied voltage is  assumed to be alternating, the frequency of the field affects the hopping rate as well. In addition, at a particular frequency, the charge carriers can hop between two localized sites back and forth which is observed as if dipoles were present~\cite{Jonscher1983,MacDonald1987}. Therefore, at low frequencies or higher temperatures, it is not easy to separate conduction and polarization mechanisms in insulators. The frequency dependent complex conductivity $\widetilde\sigma$ is given as \cite{Jonscher1983,MacDonald1987},
\begin{gather}
    \begin{split}
  \label{eq:acconductivity}
  \widetilde\sigma(\omega)&=\imath\varepsilon_0\varepsilon(\omega)\\
  &=\sigma+\varsigma(\imath \omega)^n
    \end{split}
\end{gather}
where $\varsigma$ is the hopping conductivity. The low frequency dispersion~\cite{Jonscher1983} or hopping conduction contribution, second term on the right-hand-side of Eq.~\ref{eq:acconductivity}, can also be included in the complex dielectric susceptibility $\chi$
\begin{equation}
\chi(\omega)=\frac{\varsigma}{\varepsilon_0 (\imath \omega)^\gamma}
\label{eq:sigmahop}
\end{equation}
where $\varsigma$ and $\gamma$ are constant parameters, $0 < \gamma \le 1$. The relationship between $n$ and $\gamma$ in Eq.~(\ref{eq:acconductivity}) and (\ref{eq:sigmahop}) is $\gamma=1-n$.

\subsection{Dielectric loss and relaxation in materials}\label{sec:Electr-prop-mater}
\subsubsection{Single relaxation time}
Debye~\cite{Debye1945} considered noninteracting dipoles in a viscous medium. He assumed a response function with a single relaxation time, $\tau_D$,
\begin{gather}
  {\sf f}(t)=\frac{\Delta\varepsilon}{\tau_D}\exp{\left(-\frac{t}{\tau_D}\right)}\label{eq:debyetime}
\end{gather}
The time constant $\tau_D$ in the equation is called the Debye dielectric relaxation time, or simply the relaxation time. The complex dielectric susceptibility, $\chi$, as a function of angular frequency, $\omega$, after using the Fourier transform of the response function in Eq.~\ref{eq:debyetime} is
\begin{gather}
\begin{split}
\chi(\omega)=&\frac{\Delta\varepsilon}{1+\imath \omega \tau_D}\\
           =&\frac{\Delta\varepsilon}{1+\omega^2 \tau_D^2}-\imath \frac{\Delta\varepsilon\omega \tau_D}{1+\omega^2 \tau_D^2}
\label{debyeeq}
\end{split}
\end{gather}
However, at low frequencies, losses due to the dc conduction are present, then,  the susceptibility contains additional term,
\begin{gather}
\begin{split}
\chi(\omega)=&\frac{\Delta\varepsilon}{1+\omega^2 \tau_D^2}-\imath \biggl [\frac{\Delta\varepsilon\omega \tau_D}{1+\omega^2 \tau_D^2}+\frac{\sigma}{\varepsilon_0 \omega} \biggr ]
\label{debyeeq2}
\end{split}
\end{gather}
It is possible to distinguish the two independent processes, dipolar relaxation and dc conduction using the Kramer-Kr{\"o}nig relation~\cite{LL,Tuncer2000b,Steeman1997} or using curve fitting algorithms based on complex non-linear least squares~\cite{Tuncer2000b,BoPhD,Tuncer2001a,MacDonald1987,MacDonald1986,Macdonald81,Macdonald77}.

\subsubsection{Distribution of relaxation times}
From the experimental point of view, the dielectric response of solid materials, except ferroelectrics, does not often show Debye characteristics, and their relaxation characteristics are stretched out in the time domain, and similarly in the frequency domain. For this reason, they are rarely modeled by distribution of relaxation times~\cite{Tuncer2000b,Carmona99,Kliem88,Mopsik,Yager1936} or often by a non-Debye empirical response~\cite{Tuncer2000b,Cappaccioli2000,Capaccioli1998,Scaife,Sidebottom97,Al-Refaie1996,Bo96,Phillips,Gubta1994,DasGubta2,Schlonhals1989,WilliamsIEEE,Jonscher1983,Daniel,McCrum,Kauzmann,CC,Kirkwood1941,Yager1936}. The former model considers a distribution function, $\mathcal{F}(\tau)$, for the relaxation time, $\tau$. Consequently, Eqs.~\ref{eq:debyetime} and \ref{debyeeq}, then, become,
\begin{gather}
\begin{split}
{\sf f}(t)=&\Delta\varepsilon\int_0^\infty\mathcal{F}(\tau)\exp\left ( -\frac{t}{\tau}\right)\text{d}\tau\\
\chi(\omega)=&\Delta\varepsilon\int_0^\infty\frac{\mathcal{F}(\tau)}{1+(\imath \omega \tau)}\text{d}\tau
\label{eq:drt}
\end{split}
\end{gather}
$\mathcal{F}(\tau)$ can be regarded as the fraction of polarization processes with relaxation times between $\log \tau$ and $\log \tau+\dd \log \tau$ and it satisfies the condition
\begin{equation}
  \label{eq:F(t)}
  \int_{-\infty}^\infty \mathcal{F}(\log \tau)\dd \log\tau =1
\end{equation}

There have been several methods proposed to obtain $\mathcal{F}(\tau)$~\cite{Morgan,Kremer,Keiter98,Carmona99,Kliem88,Mopsik,Yager1936,KirkFuoss,Kirkwood1941}, and, the problem is considered ill-mannered. However, using a least-squares-algorithm with constraints~\cite{Adlers} and applying the Monte Carlo technique~\cite{Binder,Gershenfeld}, a new procedure has been presented~\cite{Tuncer2000b}. In the procedure, a least-squares method with box-constraints was used~\cite{Adlers} and with the application of the Monte Carlo algorithm a continuous time-axis was constructed. The survived time constants from the least squares yielded the distribution of relaxation times.

\begin{figure}[pbt]
    \centering{\includegraphics[height=3.3in,angle=-90]{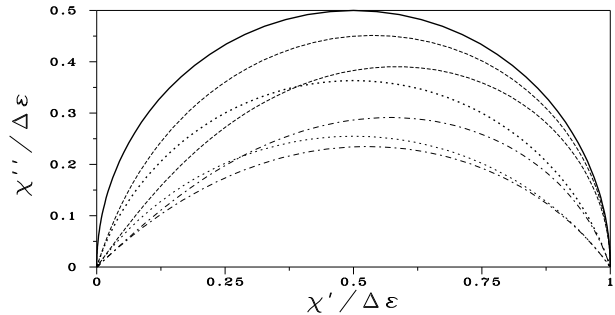}}
\caption{\label{fig:debye0}
Normalized dielectric relaxation of different functions, Debye single relaxation (\full), Cole-Cole with $\alpha=0.6$ and $\alpha=0.8$ (\dotted), Davidson-Cole with $\beta=0.6$ and $\beta=0.8$ (\dashed) and Havriliak-Negami with $(\alpha,\beta)=(0.6,0.8)$ and $(\alpha,\beta)=(0.8,0.6)$ (\longchain). 
}
\end{figure}

Since it is hard to obtain the actual distribution of relaxation times, empirical formulas in the frequency domain based on non-Debye relaxations have been presented in the literature~\cite{CC,CD,Cole55,HN,Kirkwood1941}. One of the most frequenctly used can be preseneted as:
\begin{equation}
\chi(\omega)=\frac{\Delta\varepsilon}{[1+(\imath \omega \tau)^\alpha]^\beta}
\label{eq:havneg}
\end{equation}
This equation is called the Havriliak-Negami formula~\cite{HN}, and  $\alpha$ and $\beta$  are empirical parameters depending on the shape of the response, ($0 <\alpha \le 1\quad \land \quad \alpha\beta\le1$)~\cite{Bo96}. When, $\alpha=\beta=1$, we have the classical Debye process. The other famous emprirical response functions with distributed relaxation times can also be obtained from Eq.~(\ref{eq:havneg}), {\em i.e.}, for  $\alpha=1$ and $\beta\not=1$ the Cole-Davidson~\cite{CD} and for $\alpha\not=1$ and $\beta=1$ the Cole-Cole~\cite{CC} empirical formulas. In Fig.~\ref{fig:debye0} different dielectric functions are presented as Cole-Cole plots ($\chi''(\omega)$ versus $\chi'(\omega)$).

When the conduction processes are taken into account, the complex dielectric susceptibility of a material, $\chi$, can therefore  be expressed in an empirical form as 
\begin{equation}
\chi(\omega)=\sum_j\frac{\Delta\varepsilon_j}{(1+(\imath \omega \tau_j)^{\alpha_j})^{\beta_j}}+\frac{\sigma}{\imath \varepsilon_0 \omega} + \frac{\varsigma}{\varepsilon_0 (\imath \omega)^\gamma}
\label{eq:gener}
\end{equation}
where summation over $j$ represents different dielectric relaxation processes within the material, or in a more general functional form as
\begin{equation}
\chi(\omega)=\int_0^\infty\frac{\Delta\varepsilon\mathcal{F}(\tau)}{1+(\imath \omega \tau)}\text{d}\tau+\frac{\sigma}{\imath \varepsilon_0 \omega}
\label{eq:gener2}
\end{equation}

\section{Studies of dielectric mixtures}

\subsection{Experimental studies of dielectric systems}

Techniques widely used in different branches of science  for studying dielectric properties of materials are the dielectric spectroscopies in time and frequency domains. They provide informations on structural and dynamical properties of considered systems. There exits in the literature vast of articles on the applications of the dielectric spectroscopy to mixtures.  A broad range of frequencies, $\nu$, can be accessed.  The information obtained about the relaxation rates and time constants $\tau$ (or the dynamics) of a system ranges approximately between $\micro\hertz$ to $\peta\hertz$ in the frequency domain and between $\mega\second$ to $\femto\second$ in the time domain, respectively. The polarization mechanisms, such as, electronic, atomic, molecular {\em etc.,} in the materials can be therefore studied. Moreover, the dielectric spectroscopy is an effective tool to investigate not only pure form of material states on any scale (micro-, meso- and macroscopic), but their interactions as well.  

For insulation applications polymer matrix has been filled with inorganic powders, for example silica and aluminum hydroxide, to enhance the mechanical and electrical properties. Similar systems can be found elsewhere~\cite{Lian,Khastgir,Kirst,Jeffery1995,KimCherney,YamanakaFukuda1,YamanakaFukuda2,BrosseauQue,Steeman90,Steeman91,Nagata,EvaLic,Ghany,Arous1997,YinTanaka,YehBudenstein,Wojtecki,Zukoski}. Previously it was noted that the dielectric permittity of a system is dependent on moisture. In some cases, such as insulation of power equipment and components, the moisture content in the system affects not only the overall insulation property but the lifetime of the system as well. Therefore, measurements of dielectric response can provide information on  the moisture content of insulating materials~\cite{RobertsPhD,AndersPhD,AndersLic,Uno98,Gafvert1996,Gafvert1000,Karner1984,Tuncer1998a,WangGubanski,ABBsiliconereport}. There are special cases where composites have been used for current limiting applications~\cite{DuggalSun,DuggalLevinson1,DuggalLevinson2,Strumpler,Strumpler2,Strumpler3}. In those cases knowledge of the dielectric properties is essential. In other research fields than electrical insulation, the water uptake of materials and their dielectric properties have been also studied~\cite{mattssongunnar,Zakri,Banks,Capaccioli3,Capaccioli2,BergNiklasson,SheiretovZhan,Varlow,Kabir,Jeon}. 
Liquid-solid mixtures~\cite{Brosseau}, such as, brine-saturated rocks and porous media~\cite{Yoon,BoJMS,Bo1,Bo96,BoPhD,Bo2,Capaccioli3,Capaccioli2,Haslund}, suspensions in solutions~\cite{Ishida,Arroyo,PS99} or liquid crystal in polymer matrices~\cite{Boersma1,Boersma2,Reshetnyak,Briganti,Aboudi} have been examined using dielectric spectroscopy technique. 
One can also include the mixture examples in biological systems where the dielectric spectroscopy has been one of the tools to obtain information about the structural and dynamical behavior~\cite{Brown,Polevaya,Feldman1,Raicu}. Similarly, work on liquid-liquid mixtures can be found in the literature~\cite{SmithLee,Henderson,WangWang,Bertolini,McAnanama,Suzuki,DuttaBasak,Mixture1,Smith2,Hanna1,UrbanGest,CooperHill,BeckerStock,Jellema,Skodvin,Grosse1,Grosse2}. The dielectric spectroscopy technique has been found powerful for investigating polymer-polymer mixtures~\cite{Bhattacharyya,Dutta,ZhangYi,Nuriel,Becker,Factor,Snoopy1,Boersma3,Hayward,ChanZhang,YoshinoYin}. Recently, dielectric behavior of confined polymers has got attention~\cite{ChoWatanabe,Pelster1,Arndt1,Kirst,Svanberg}. In these cases, guest molecules are trapped in regular matrices of polymers or on substrates. Such structures can also be regarded as composites.

Charge relaxation phenomena in mixtures composed of ceramics~\cite{Benziada,Mururgan,Ragossnig,Alderbas}, metal-oxides~\cite{Murawski1,Murawski2}, ionic-conductors~\cite{Debierre,Siekierski,Cappaccioli2000,Henn,Tengroth} have been also studied~\cite{Murawski1,Benziada,Murawski2,Mururgan,Ragossnig,Lian,Alderbas}. In the case of conductive coating and   electromagnetic shielding applications carbon-black~\cite{Park,YuZhang,ZhangYi,Kupke1998,Ziembik,Calberg,Ramos2,BrosseauBoulic,Chellappa,Ardi} and metallic particles~\cite{Uberall,Craig2,BrosseauQue} have been used as heterogeneities in the matrix medium. 
For photonic applications, metal-dielectric mixtures have been widely applied~\cite{Cummings1984,Tsangaris1996,Tsangaris1999,Neelakanta,abe75,she84,Pike,Ottavi,Tesfamichael,Matsumoto,ban98} to obtain photon localization.

Finally, dielectric spectroscopy or similar techniques has been used to sense the composition of structures~\cite{Ramos,Huges,Shkel,Asami,Mixture2,Mixture3,Strumpler3,LiuRosidian} or even buried objects~\cite{Osella,Lattarulo,Reppert,Saarenketo,Olhoeft}. Experimental studies have shown that mixtures have enormous potential for special applications. Therefore, numerical simulations of mixtures can lead not only better understanding of the physics of dielectrics but also improvement of designing of tailored materials without going to expensive attempts.

\subsection{Different analytical approaches}

Predicting the dielectric properties of mixtures has been a challenging problem when structures other than the layered arrangements were considered. These different approaches are reviewed below.

\subsubsection{The simplest approach}
Gladstone and Dale~\cite{AragoBiot} expressed a formula for the effective electrical properties of mixtures which was proportional and linear to the concentrations and dielectric permittivities of constituents
\begin{equation}
  \label{eq:simplest}
  \widetilde\varepsilon_{\sf e}=(1-q) \widetilde\varepsilon_1 +q \widetilde\varepsilon_2
\end{equation}
In reallity, all the expressions that are presented later should yield the same value as Eq.~(\ref{eq:simplest}) for very low additive concentration ($q\rightarrow0$)~\cite{Lakhtakia1}, if the ratio of the complex premittivities is not too large.
\subsubsection{Mean field theories}

The mean field theory approaches suppose that one of the phases, an inclusion phase, is inside a matrix phase, and both phases are embedded in an effective medium. Wagner~\cite{Wagner1914} was the first to use this approach, in which in the same way as Maxwell~\cite{Maxwellbook}, he assumed a structure composed of an insulating host medium, $\widetilde\varepsilon_1=\epsilon_1$, with dispersed conducting spherical particles with complex permittivity, $\widetilde\varepsilon_{2}=\epsilon_2+\sigma_2/(\imath\varepsilon_0\omega)$. The volume fraction of the spherical inclusions was $q$. 
Later, Sillars~\cite{Sillars1937} and Fricke~\cite{Fricke1924} expanded the considerations for conducting ellipsoidal particles dispersed in an insulating medium~\cite{VanBeek1960}. In this case shape of the inclusions was accounted for by the parameter $n$ in the complex dielectric permittivity of the mixture.
The shape factor $n$ is a function of the ellipsoidal axes and of the desired direction~\cite{Sillars1937}:
\begin{gather}
\begin{split}
n_i & =\frac{2}{x_1 x_2 x_3 \mathcal{L}_i}
\label{eq:shapefactorn}
\end{split}
\end{gather}
where the subscript $i$ is the desired axis-direction, $i=\{1,2,3\}$. The vectors $x_{1,2,3}$ are the radii of the ellipsoid in the arbitrary directions $1$, $2$ and $3$, and $\mathcal{L}_i$ is 
\begin{gather}
\begin{split}
\mathcal{L}_i=\int_0^{\infty}\frac{d\zeta}{\biggl[ (x_i^2+\zeta)^2 (x_1^2+\zeta)(x_2^2+\zeta)(x_3^2+\zeta) \biggr ]^{1/2}}
\label{eq:shapefactorL}
\end{split}
\end{gather}
where $\zeta$ is the integration variable. The value of $\mathcal{L}_i$ is between 1 and infinity, $1<\mathcal{L}_i<\infty$. 
For needle-like shapes parallel to the field direction, $n$-values are larger than 3. For  spherical inclusions, the shape factor, $n_i$, is 3, and for cylindrical inclusions if the cylinder axis is perpendicular to the applied field, the shape factor is 2, and if large disks perpendicular to the applied field are considered, the shape factor is 1. 
As a matter of fact, this ellipsoidal model for the dispersed particles represents an infinite variety of shapes which  describes real composite materials well. Steeman and his collaborators~\cite{Steeman90,Steeman1992} have improved the model of interfacial polarization of two phase systems by considering the case when the host medium was not insulating but conductive, which was also mentioned elsewhere~\cite{Weber,Sihvola-ellips,Sihvola-sphere}. 
The effective permittivity of the mixture was expressed as in the form of Eq.~(\ref{eq:maxw1})~\cite{Steeman1992}, with
\begin{gather}  \begin{split}
    \widetilde\varepsilon_{\sf e}(\omega)=&\frac{\widetilde\varepsilon_{1} [ \eta \widetilde\varepsilon_2 + (1-\eta) \widetilde\varepsilon_1] - q (1-\eta) (\widetilde\varepsilon_2 -\widetilde\varepsilon_1 )}{[\eta\widetilde\varepsilon_2+(1-\eta)\widetilde\varepsilon_1]-\eta q (\widetilde\varepsilon_2-\widetilde\varepsilon_1)}\\
    \Delta\varepsilon=&\frac{\eta q (1-q)(\widetilde\varepsilon_2 \sigma_1-\widetilde\varepsilon_1\sigma_2)^2}{[(1-\eta)\widetilde\varepsilon_1+\eta\widetilde\varepsilon_2+\eta q (\widetilde\varepsilon_1-\widetilde\varepsilon_2)][(1-\eta)\sigma_1+\eta\sigma_2+\eta q (\sigma_1-\sigma_2)]^2}\\
    \tau=&\varepsilon_0 \frac{(1-\eta)\widetilde\varepsilon_1+\eta\widetilde\varepsilon_2+\eta q (\widetilde\varepsilon_1-\widetilde\varepsilon_2)}{(1-\eta)\sigma_1+\eta\sigma_2+\eta q (\sigma_1-\sigma_2)}\label{eq:steeman}
\end{split}\end{gather}
where $\eta$ is the inverse shape factor, $\eta=n^{-1}$. Note that for $\eta=1$ the equations above yield epxressions for Maxwell's laminated structure, Eq.~(\ref{eq:Maxw}). Similarly, Wagner's~\cite{Wagner1914} expression can be obtained by considering $\eta=1/3$. 
Steeman and Maurer~\cite{Steeman90} have also introduced a third phase to the system where the third phase is an outer shell of the second phase (phase 3). 

The mean field approximations are valid for low concentration of inclusions ($q\ll1$), since in such cases the inclusion particles are apart from each other and can not feel the influence of the closest neighbors -- dipole-dipole interactions~\cite{Emets_inter,eme96,Lopez,Bozdemir1,Bozdemir2,Tokuyama,Toriyama,kel96,Keller2} which accordingly are ignored. 

\subsubsection{Molecular approaches}

In these approaches, the dielectric behavior of the mixture is assumed to be the summation of the dipole moments of each molecule in a vacuum matrix. Similarly to the mean field approach each molecule feels only the force of the applied field, not the neighboring molecules -- no dipole-dipole interactions are included. Examples to the molecular approaches are Clausius-Mosotti~\cite{Bottcher,SihvolaBook,greffePier6}, Onsager-B{\"o}ttcher~\cite{Onsager,Bottcher,SihvolaBook,greffePier6}, and  the effective permittivities are expressed as in Eqs.~(\ref{eq:maxw2}) and (\ref{eq:steeman}). 
Kirkwood~\cite{Kirkwood} extended the Onsager theory of dielectric polarization by including the influence of nearest neighbors.

\subsubsection{Regular arrangement of inclusions}

Rayleigh~\cite{Rayleigh} approached to the binary mixture problem from a different point of view by assuming unit cells. 
In his approach, repeating unit cells composed of a matrix phase with sphere inclusions in the centers were considered. 
He assumed that both media were dielectrics without any conductivity, then, the permittivity of the composite was expressed analytically  as a series: 
\begin{gather}
  \begin{split}
    \widetilde\varepsilon=\widetilde\varepsilon_{1}\left[\frac{\widetilde\varepsilon_{2}+2\widetilde\varepsilon_{1}+2q(\widetilde\varepsilon_{2}-\widetilde\varepsilon_{1})-q^{10/3}\frac{\pi^2(\widetilde\varepsilon_{2}-\widetilde\varepsilon_{1})^2}{6(\widetilde\varepsilon_{2}+4\widetilde\varepsilon_{1}/3)}+\dots}{\widetilde\varepsilon_{2}+2\widetilde\varepsilon_{1}-q(\widetilde\varepsilon_{2}-\widetilde\varepsilon_{1})-q^{10/3}\frac{\pi^2(\widetilde\varepsilon_{2}-\widetilde\varepsilon_{1})^2}{6(\widetilde\varepsilon_{2}+4\widetilde\varepsilon_{1}/3)}+\dots}\right]
  \label{eq:rayleigh}
  \end{split}
\end{gather}
Rayleigh's expression has also been derived by Lorentz--Lorentz~\cite{Bottcher,Lakhtakia1} and Maxwell-Garnett~\cite{Maxwell_Garnett}.  The model includes the interaction between spherical inclusions, therefore, by contrast to Wagner~\cite{Wagner1914} and Sillars~\cite{Sillars1937}, it should be applicable to mixtures with higher concentration of inclusions. However, when the concentration of the inclusions is approaching one, $q\rightarrow 1$, the permittivity of the mixture, $\widetilde\varepsilon$ is according to Eq.~(\ref{eq:rayleigh}) not equal to the permittivity of the inclusions, $\widetilde\varepsilon_{2}$. Although the model can be used for irregular distributions of dilute mixtures, 
 for higher concentrations of inclusions this formula deviates from reality~\cite{McPhedran2,doy78}. 
Bruggeman~\cite{Bruggeman1935} used Eq.~(\ref{eq:rayleigh}),  and introduced the incremented volume fraction, $\text{d}q$, of the inclusion phase. He has then derived the following expression:
\begin{gather}
  \begin{split}
    (1-q)\frac{\widetilde\varepsilon_{1}-\widetilde\varepsilon_{\sf e}}{\widetilde\varepsilon_{1}+2\widetilde\varepsilon_{\sf e}}+q\frac{\widetilde\varepsilon_{2} -\widetilde\varepsilon_{\sf e}}{\widetilde\varepsilon_{2}+2\widetilde\varepsilon_{\sf e}}=0
  \end{split}
\label{eq:bruggeman0}
\end{gather}
This equation is known as the symmetric Bruggeman formula. The non-symmetric Bruggeman formula is~\cite{Bruggeman1935,Hanai1960},
\begin{gather}
  \begin{split}
    \frac{\widetilde\varepsilon_{2}-\widetilde\varepsilon_{\sf e}}{\widetilde\varepsilon_{2}-\widetilde\varepsilon_{1}}=(1-q)\left(\frac{\widetilde\varepsilon_{\sf e}}{\widetilde\varepsilon_{\sf 1}}\right)^{\eta}
  \end{split}
\label{eq:bruggeman1}
\end{gather}
Here again $\eta$ can be regarded as the inverse shape factor as in Eq.~(\ref{eq:shapefactorn}) or space dimensionality~\cite{SihvolaBook,Landauer1978}. 

At higher filler concentrations the mutual interaction of particles is important. Interactions between two particles with different diameters and  different dielectric properties which were placed in a background medium were presented by Emets~\cite{eme96} in two-dimensions. The mutual influence of neighboring particles, {\em e.g.,} force, $F$, not only depends on the mentioned properties, but it is also affected by the direction of the applied electric field. The polarization of each particle in the applied field affects the field distribution, and therefore, the polarization of the neighboring inclusions~\cite{kel96}. 

The Rayleigh model was improved by Emets~\cite{Emets98b} for systems with three components which had complex dielectric permittivites, $\widetilde\varepsilon_{i}\ne\epsilon_{i}$. In this case, the inclusions were assumed to be doubly periodic in an alternating order, and these were two different cylinders with radii, $r_1$ and $r_2$. The complex effective dielectric constant of the system, $\widetilde\varepsilon$, was calculated from the averages of the displacement and electric field vectors in the region of interest, 
  \begin{gather}
    \begin{split}
      \widetilde\varepsilon=&\widetilde\varepsilon_1\frac{\mathcal{A}}{\mathcal{B}}\\
       \mathcal{A}=&1-\Delta_{12}q_{1}/2-\Delta_{13}q_{2}/2+\Delta_{12}^2A_1
         +\Delta_{13}^2A_2+\Delta_{12}\Delta_{13}(B_1+B_2)\\
       \mathcal{B}=&1+\Delta_{12}q_{1}/2+\Delta_{13}q_{2}/2+\Delta_{12}^2A_1            +\Delta_{13}^2A_2+\Delta_{12}\Delta_{13}(B_1+B_2)
      \label{eq:emets}
    \end{split}
  \end{gather}
Here
\begin{equation}
  \label{eq:delta}
  \Delta_{1p}=\frac{\widetilde\varepsilon_1-\widetilde\varepsilon_p}{\widetilde\varepsilon_1+\widetilde\varepsilon_p} \qquad (-1\le\Delta_{1p}\le1)\qquad p=2,3
\end{equation}
where, $q_{1,2}$ are the concentrations of the inclusion phases, ($0\le q_{1,2}\le 1$). 
The parameters $A_k$ and $B_k$ ($k=1,2$) are functions of the radii of the inclusions,
\begin{gather}
  \begin{split}
    A_k=&2r_k^2 \Biggl\{
        2r_k^3\sum_{m=1}^{\infty} \frac{1}{r_k^4-16m^4}
        +\sum_{n=1}^{\infty}\sum_{m=1}^{\infty} \biggl[
        \frac{r_k-2m}{(r_k-2m)^2-4n^2}\\
        &+\frac{r_k+2m}{(r_k+2m)^2-4n^2}+\frac{r_k-2m+1}{(r_k-2m+1)^2-(2n-1)^2}\\
&+\frac{r_k+2m-1}{(r_k+2m-1)^2-(2n-1)^2}\biggr] \Biggr\}
  \end{split}
\label{eq:emets3_1}
\end{gather}
\begin{gather}
  \begin{split}
    B_k=&2r_{3-k}^2 \Biggl\{
        2r_k^3\sum_{m=1}^{\infty} \frac{1}{r_k^4-(2m-1)^4}
        +\sum_{n=1}^{\infty}\sum_{m=1}^{\infty} \biggl[\frac{r_k-2m}{(r_k-2m)^2-(n-1)^2}\\
        &+\frac{r_k+2m}{(r_k+2m)^2-(n-1)^2}+\frac{r_k-2m+1}{(r_k-2m+1)^2-4n^2}\\
        &+\frac{r_k+2m-1}{(r_k+2m-1)^2-4n^2} \biggr] \Biggr\}
  \end{split}
\label{eq:emets3_2}
\end{gather}
$A_k$ and $B_k$ values converge quickly to a constant level for $n=m\ge10$. When the Rayleigh conditions are applied, Eq.~(\ref{eq:emets}) becomes equal to Eq.~(\ref{eq:rayleigh}). Although these equations for regular structures with round shapes (spheres or cylinders) are valid for low concentrations, at high frequencies the sizes of the particles are playing an important role~\cite{Kottmann,PhillipsPRB,FuchsLiu,Cummings1984,Datta,Malliaris1971,Stoyanov}. Finally, other periodic structures have also been presented in the literature such as spherical particles~\cite{Taniguchi,Blase,doy78,McPhedran2,McPhedran1}, cylindrical inclusions~\cite{mil81,Emets98b} tilings (checkerboard) structures~\cite{Helsing1,EmetsObnosov2,EmetsObnosov1,Emets_sym,Ke-da,HuiKe-da} and layered structures~\cite{Mahan,Tretyakov,Vasetskii}.

\subsubsection{The spectral function approach}
A theoretical advance to the modeling of dielectric  properties of binary composites has been developed independently by Bergman~\cite{Bergman1979,Bergman1978,Bergman3,Bergman4} and Milton~\cite{Miltona,mil81,Milton1981b,Milton1981,Stroud}. The method is called the spectral theory, in which a function of composite has been introduced as a compact way of representing data over a range of frequencies and it highlights the role of geometry in determining the effective properties~\cite{Ghosh,GhoshFuchs,Day}. In this model, the dielectric permittivity of the composite $\widetilde\varepsilon_{\sf e}$ is expressed as a function of dielectric permittivities of the constituents $\widetilde\varepsilon_1$ and $\widetilde\varepsilon_2$ and the geometry of the composite. 

\subsection{Bounds on electrical properties}
Difficulties in calculating the effective electrical properties of mixtures lead to interest in obtaining bounds on these parameters. Wiener~\cite{wiener} and Hashin-{Shtrik\-man}~\cite{hashin} proposed bounds for two component systems. Wiener assumed layered structures which had topological configurations parallel or perpendicular to the applied field direction. According to it
\begin{equation}
  \label{eq:wiener}
  \widetilde\varepsilon_1\widetilde\varepsilon_2 [q\widetilde\varepsilon_1+(1-q)\widetilde\varepsilon_2]\le \widetilde\varepsilon_{\sf e}\le(1-q)\widetilde\varepsilon_1+q\widetilde\varepsilon_2
\end{equation}
These bounds are called absolute bounds meaning that the effective property, $\widetilde\varepsilon_{\sf e}$ can not be present outside the region given in Eq.~(\ref{eq:wiener}). If one of the phases forms inclusions (phase 2) in the other (phase 1), then, these two expressions can further be put together by introducing a factor $n$~\cite{greffePier6}, which can be treated as the shape factor of the inclusion phase.
\begin{equation}
  \label{eq:wiener_pier}
  \frac{\widetilde\varepsilon_{\sf e}-\widetilde\varepsilon_{1}}{\widetilde\varepsilon_{\sf e}+(n-1)\widetilde\varepsilon_{1}}=q\frac{\widetilde\varepsilon_{2}-\widetilde\varepsilon_{1}}{\widetilde\varepsilon_{2}+(n-1)\widetilde\varepsilon_{1}}
\end{equation}
Eq.~(\ref{eq:wiener_pier}) reduces to Wiener bounds as expressed in Eq.~(\ref{eq:wiener}) for $n=1$ and for $n=\infty$. Moreover, Eq.~(\ref{eq:wiener_pier}) is also the same as the Rayleigh~\cite{Rayleigh} and Maxwell-Garnett~\cite{Maxwell_Garnett} expressions, Eq.~(\ref{eq:rayleigh}). Landauer~\cite{Landauer1978} has mentioned that $n$ is the space dimensionality, where $n=2$ or $n=3$. His approach may be valid, as in other mixture formulas, {\em i.e.,} the Sillars derivation using the effective medium theory for a composite with arbitrary shapes~\cite{Sillars1937} or the Bruggeman assymetrical theory~\cite{Bruggeman1935} using the differential effective medium theory, the $n$-value represents the shape of the mono-dispersed inclusion particles. 

Hashin-Shtrikman bounds are narrower than Wiener bounds and they are derived for models of homogeneous and isotropic mixtures. In two-dimensions, the Hashin-Shtrikman bounds are expressed as~\cite{SihvolaBook}
\begin{equation}
  \label{eq:hashin}
  \widetilde\varepsilon_1+\frac{q}{(\widetilde\varepsilon_2-\widetilde\varepsilon_1)^{-1}+(1-q)(2\widetilde\varepsilon_1)^{-1}} \le \widetilde\varepsilon\le \widetilde\varepsilon_2+\frac{1-q}{(\widetilde\varepsilon_1-\widetilde\varepsilon_2)^{-1}+q(2\widetilde\varepsilon_2)^{-1}}
\end{equation}
Other bounds have been proposed by Bergman~\cite{bergman1982,Bergman3,Bergman1978}, Milton~\cite{Milton1981,Milton1981b,Miltona} and Golden and his collaborators~\cite{Golden3,Golden2,Golden4,Golden1}. It was found that (see Tuncer et al.~\cite{Tuncer2001b}) the bounds were not valid for low frequency permittivity values when interfacial polarization was considered.

\subsection{Percolation theory}

When higher inclusion concentrations or higher ratios of electrical properties of constituents~\cite{HuiKe-da} are  considered, percolation phenomenon should be taken into account~\cite{Clerc1990,Shalaev1996,Clerc1996,Kirkpatrick1,Kirkpatrick2,Kirkpatrick3}. It is therefore more meaningful to speak of percolation in disordered systems~\cite{bid93}. At some specific concentration, depending on the model of the packing density, particles start to form conducting chains (percolating clusters) which influence the electric properties, the overall conductivity and the dielectric relaxation~\cite{Clerc1990}. In regularly distributed systems composed of hard spheres (in 3-dimensions) or hard disks (in 2-dimensions) percolation is only observed close to limiting concentrations~\cite{McPhedran1,McPhedran2,TuncerPhD,Tuncer1998b}. Other studies have shown that not only the percolation of one of the phases but the field, charge~\cite{Shalaev1996,kel96,fer90,Mott,Zekri} and particle related issues~\cite{ban98,YuZhang,Wang1993,JingZhao,EvaLic,Garncarek} should also be considered.

In the case of disordered systems, their physical behavior can usually be described by a power law with a percolation threshold and a critical exponent depending on the geometry~\cite{Turban,Sheng,Ottavi,Ottavi2} and on the  conduction of the system~\cite{Clerc1990,Shalaev1996,Kirkpatrick2,Kirkpatrick1,Deutscher}. One should also not exclude the local field changes in the composite due to interaction of inclusions. Sorbella~\cite{sor84} concluded that the conductivity of the system around the localized scatterers (inclusions) did not obey Ohm's law. This shows that the micro-structure of a heterogeneous medium affects the conductivity of the system depending on a length scale, $l_{c}$ which was defined as the distance between the scatterers or inclusions. Mendelson and Schwacher~\cite{men84} have also mentioned a similar length scale, $l_{d}$, which is the variation of the dielectric permittivity in a continuous random dielectric. They have emphasized an other length scale, $L_{q}$; variation of charge density or the electric field. Therefore, average values for the physical quantities such as electric field, potential, dielectric permittivity etc. can be used when $L_{q}\gg l_{d}$ or $L_{q}\gg l_{c}$.

High concentration of inclusions in composite structures brings forth the problem of particle-particle interactions~\cite{Kirkwood,Huang,Toriyama,Emets_inter,eme96}. The question of packing density for the particulated composite materials~\cite{Kauly1997,Tsirel1997} and of the packing structure itself becomes important. Particle size, shape, size distribution and cohesion force between them affect the maximum filler content~\cite{Malliaris1971} and isotropy of the material.  In the percolation region (concentration) and over, the theories described in this section become unreliable. However, some of the mixture formula might be used to predict the percolation threshold, in such cases the shape parameter $n$ in the mixture formulas may be connected with the threshold~\cite{SihvolaBook}.

The electric conductivity and its mechanisms in granular materials containing metallic inclusions  have been investigated intensively~\cite{fer90,Ottavi,abe75,coh73,Abeles,Pike}. Cases when the concentration of the metal inclusions were low enough were interesting~\cite{coh73,abe75,she84,Abeles} due to possible technological applications. However, it has not been easy to estimate metal concentration for the metal-insulator transition~\cite{Percolation,Kirkpatrick1,Landauer1952,Cohen}, since the tunneling probabilities between the metal inclusions (or the more conductive phase) are involved~\cite{fer90,Abeles1}.  For low concenrations of inclusions, in which the system is under the percolation threshold, the conductivity can be activated by the tunneling process. 
When the percolation threshold is passed, the system clearly shows dc conductivity which is similar to transition between hopping and dc conductivity~\cite{Mott,Zallen,Elliot_Amorphous}.

\section{Numerical simulations}
\subsection{Previous approaches}
Numerical simulations on the behavior of composites can be performed using different techniques.  One of the ways is to solve Maxwell's equations in a computational domain~\cite{Booton,Silvester1990,Mackerle,Hubing} and to obtain either the electrical potential or electric field distributions in the domain. However, differences in the physical dimensions in the domain should not be too large, which is one of the drawbacks of computer simulations. Once the distribution of these quantities are known the effective material parameters can be calculated using different methods~\cite{Tuncer-CEIDP01}.  The finite element method ({\sc fem}) is one of the most intensively used approaches in the field simulations~\cite{Booton,Silvester1990,FEMWE}. In this study, the effective medium properties of mixtures are calculated using the finite element method~\cite{Tuncer-CEIDP01,Tuncer2001a,Tuncer2001b,Tuncer2001c1,Tuncer2001d,Tuncer2001elips,TuncerAcc1,TuncerPhD,Tuncer1999a,Tuncer1998b}. Similar approaches can also be found elsewhere~\cite{Boudida,Boudida2,Brosseau1,Brosseau2,Sareni1,sar97,sar97mag,Sareni1996,Clauzon,gho94,Sihvola-ellips,Sihvola-sphere,SihvolaEff,Tsangaris1999}. 

There is a multipole expansion approach~\cite{Emets98b,eme96,Emets_inter,Hinsen,Smith,Fu1,Fu2,Harfield} in which the considered inclusion phase is replaced by multipole source distributions. Monte Carlo techniques have been introduced in which many possible geometrical arrangements have been considered~\cite{Barrowes,Denk,LiuMC}. The Monte Carlo approaches are similar to molecular dynamic simulations. Since each medium can be characterized with its resitance, capacitance and inductance, modeling composite media with resistor-capacitor networks has studied~\cite{Dyre,Dyre_a,Dissado1,IshidaHill,Basu,Vainas,Percolation,Kirkpatrick1,Borcea,Borcea2}. In addition, if there exists a periodicity in the considered geometries, {\em i.e.,} arrangement of inclusions, fast-fourier-transformation techniques has been applied to obtain the response of the mixtures~\cite{BoPhD,BoJMS,Bo1,Bo2,Sheng1,CLiu,Bergman1,shen1990,Tao,EyreMilton}. Finally, novel approaches, like expanding the potential in harmonic functions~\cite{HelsingGrimvall,Helsing4,Helsing1,Helsing2,Helsing3}, using Gaussian random fields~\cite{Berk,Wang,Roberts1,Roberts2,Roberts3,Bergman1} and many other methods~\cite{Wang,Karafyllis,Schwartz,Torquato,Michiels,SihvolaPol,Lakhtakia3,Henderson,Dasgupta,Vila1992,Reshetnyak,Craig1,BrosseauMueller,Roussel} have been introduced. 

At high frequencies or in structures where the smallest part of inhomogeneities -- which can be regared as the size of inclusions -- are larger than the wavelength of the applied field, the effective permittivity calculations are not trivial and the results can be misleading~\cite{Kottmann,Fuchs_a,Rojas}. 

Numerous simulations confirmed the validity of the mixture formulas for regular structures when the mixtures were dilute~\cite{Tuncer2001a,Tuncer2001elips,TuncerAcc1,Tuncer1999a,Tuncer1998b,sar97,sar97mag,Sareni1996,Boudida,Sihvola-sphere,gho94,Lakhtakia3}. For dilute random mixture the results were also similar~\cite{TuncerRandom,sar97,Pier6} emphasizing the validity of the effective medium approximations at low inclusion concentrations. When higher concentration of inclusions were studied the validity of the analytical solutions and their comparisons to numerical results were usually questioned and the results were compared with those of the bounds on the electromagnetic properties of composites~\cite{wiener,hashin,Golden3,HelsingBou,Golden2,Golden4,Golden1,BergmanBou,bergman1982,Milton1981,Milton1981b,Miltona,Bergman3,Bergman4,Bergman1978,Goncharenko,Merrill}. In addition, at higher concentrations, approaches like fundamental geometrical parameters have been introduced~\cite{Miltona,Stroud,Bergman1,bergman1982,Bergman3,Bergman4}. Although the geometrical parameters of a structure were constant, one should not disregard the effect of the ratio of electrical properties of the constituens~\cite{Borcea,Borcea2,WebmanCohen} as mentioned previously.  Finally, in the numerical calculation, little attention has been given to the interfacial problem~\cite{Laskey,ChengGreengard}.

\subsection{Application of the finite element method}\label{sec:appl-finite-elem}

Analytical calculations of electromagnetic problems are limited to geometrical constraints. For some simple geometries with a small number of materials (regions) and symmetries, analytical solutions can be found~\cite{Weber,eme96,Emets98b,Jackson1975,Bottcher}. These analytical solutions are obtained using methods of images~\cite{Elliot,Eyges}, orthogonal functions (Green functions)~\cite{Steiner} and complex variable techniques~(conformal mapping)~\cite{Ramo,Eyges,Weber}. 
For more complex geometries and for non-homogeneous regions composed of several materials, numerical solutions of partial differential equations and of integral equations have been developed~\cite{Booton}. 

Numerical solutions of electrostatic problems within a non-conducting medium are based on solving Poisson's equation 
\begin{equation}
  \label{eq:Poisson}
  {\mathbf \nabla} \cdot (\epsilon\varepsilon_0 {\mathbf \nabla} \phi) = -\rho
\end{equation}
where $\phi$ and  $\rho$ denote the electrical potential and the charge density in the considered region, respectively. If the medium is conductive where no free charges and sources of charges are allowed, then, the solution is given by
\begin{equation}
  \label{eq:Laplace}
  {\mathbf \nabla} \cdot (\sigma {\mathbf \nabla} \phi) = 0
\end{equation}
When the medium is a mixture of these two cases (lossy dielectric), it consists of dielectric and conductive components. The solution, then, becomes time dependent and is given by a complex electric potential in the region with the coupling of Eqs.\ (\ref{eq:Poisson}) and (\ref{eq:Laplace}), which is also known as the continuity equation for the current density 
\begin{equation}
    {\mathbf \nabla} \cdot \left(\sigma {\mathbf \nabla} \phi \right) + \frac{\partial}{\partial t} \left \{{\mathbf \nabla} \cdot \left[ \left(\epsilon\varepsilon_0 {\mathbf \nabla} \phi \right)\right]\right\} = 0
    \label{eq:contin.1}
\end{equation}
Equivalently in Fourier-space with frequency dependent properties, {\em i.e.,} the complex dielectric permittivity, $\varepsilon(\omega)=\epsilon+\chi(\omega)+\sigma/(\imath\varepsilon_0\omega)$, where no free charges are allowed in the region, due to conductivity of the medium (lossy dielectric)
\begin{equation}
    {\mathbf \nabla} \cdot \left\{\left[ \imath \varepsilon_0\varepsilon(\omega)\omega \right] {\mathbf \nabla} \phi \right\} = 0
    \label{eq:contin.2}
\end{equation}
Field calculation softwares, such as {\sc ace} and {\sc Femlab}~\cite{AceManual,femlab}, can be used to solve Eq.~(\ref{eq:contin.2}). Once the potential distribution is known, the complex permittivity of a heterogeneous medium can be calculated in several ways, {\em e.g.}, 
\begin{enumerate}
\item[(1)] by using the total current density, $j$, and the phase difference, $\theta$~\cite{vonHippel,Scaife,Tuncer2001a,Tuncer2001b},
\item[(2)] Gauss' law and losses~\cite{Scaife,Sareni1996,Tuncer1998b,Tuncer2001b},
\item[(3)] energy balance~\cite{Tuncer2001b},
\item[(4)] and by using the average values of dielectric displacement $\langle{\mathbf{D}}\rangle$ and electric field $\langle{\mathbf{E}}\rangle$~\cite{Scaife,LL,Tuncer2001b}.
\end{enumerate}

\begin{figure}[tbp]
  \begin{center}
    \psfragscanon
    \psfrag{r3}[][]{$r_3$}
    \psfrag{r4}[][]{$r_4$}
    \psfrag{r5}[][]{$r_5$}
    \psfrag{r0}[][]{$r_{\infty}$}
    \psfrag{n3}[][]{$n=3$}
    \psfrag{n4}[][]{$n=4$}
    \psfrag{n5}[][]{$n=5$}
    \psfrag{n0}[][]{$n={\infty}$}
    \rotatebox{0}{\includegraphics[width=3in]{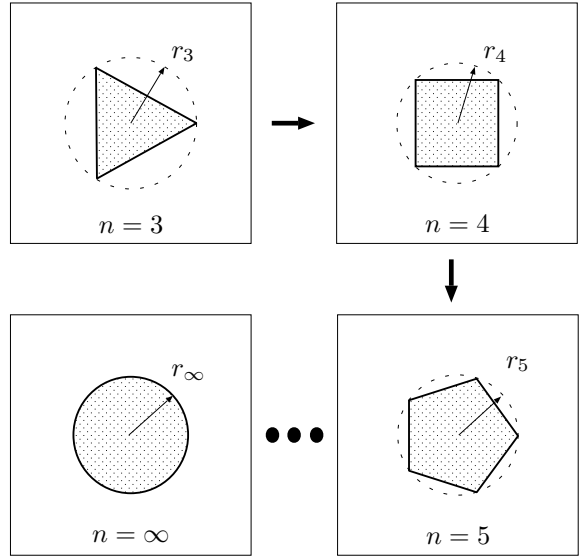}}
    \psfragscanoff
  \end{center}
  \caption{Finite-size scaling of the inclusion shapes (polygons).\label{fig:Shapes}}
\end{figure}

To check the accuracy of the finite element results, two-dimensional inclusions in the form of {\em regular} polygons with $n$ sides were considered~\cite{TuncerAcc1}, as displayed in Fig.~\ref{fig:Shapes}. The phase parameters, $\varepsilon$ and $\sigma$, were frequency and voltage independent. The values of $\varepsilon$ and $\sigma$ were chosen such that the interfacial polarization observed had a relaxation time, $\tau$, around $1\ \second$. This was achieved when the matrix phase had $\varepsilon_1\equiv\epsilon_1=2$ and $\sigma_1=1\ \pico\siemens\per\meter$, and the inclusion phase had $\varepsilon_2\equiv\epsilon_2=10$ and $\sigma_2=100\ \pico\siemens\per\meter$. The polygons were generated using a circle with radius, $r$, and a constraint on the area of the polygons, $q$. Then, the radius, $r$, as a function of $n$ and $q$ is expressed as,
\begin{equation}
  \label{eq:radius}
  r_n=\sqrt{q \over n \sin(\pi/n)\cos(\pi/n)}
\end{equation}
The denominator inside the square root approaches $\pi$ as $n\rightarrow\infty$. The size of radius was also used for the meshing procedure of the computation domain where $r_n/15$ was the size of the minimum triangle. Furthermore, this approach led to a finite-size scaling~\cite{Plischke1994,finite-size} in which as $n\rightarrow\infty$, the inclusion phase was a perfect disk. 
\begin{figure}[t]
\centering{\includegraphics[width=3in,angle=-90]{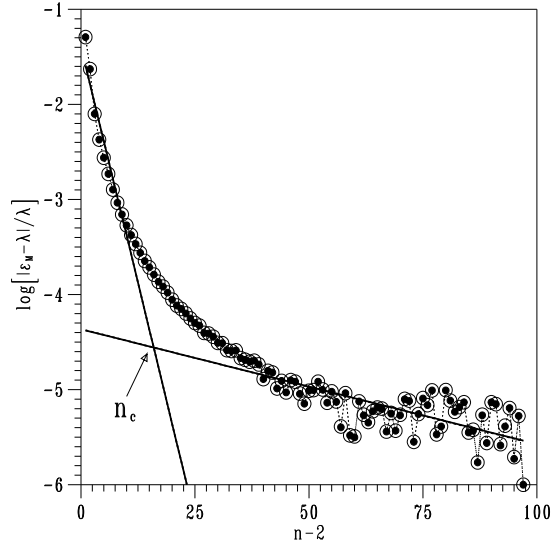}}
\caption{Dependence of percentage error in high frequency dielectric permittivity, $|\varepsilon_{\mega}-\lambda|/\lambda$ on number of regular polygon sides, $n$. The symbols open ($\bigcirc$) and filled ($\bullet$) indicate the solutions obtained using the current density and phase shift between the applied voltage and current and using Gauss' law and the total losses in the medium, respectively. The solid lines ($\full$) represents the fitted curves. $n_c$ is the critical side number for regular polygons. \label{fig:fittings4}}
\end{figure}
In Fig.~\ref{fig:fittings4}, an example of the finite-size behavior is shown. In the figure, difference between the dielectric constant at  high frequencies $\varepsilon_\mega=\varepsilon(\omega@1\mega\hertz)$ and the value obtained from analytical solution in Eq.~(\ref{eq:emets}) which is presented as $\lambda$, ($\lambda=\varepsilon_{Eq.~(\ref{eq:emets})}(\omega@1\mega\hertz)$), is plotted against the number of sides, $n$. A critical number of sides, $n_c$, is defined, such that over this value, $n>n_c$, the effective properties of medium with regular polygons as inclusions are approximately similar to those of a medium with disk shape inclusions. The analysis showed that $n_c$ is approximately $15$ regardless the concentration levels considered, $q\le0.3$, and the error in the calculations is $<0.01\ \%$ for $n_c>15$, as displayed in Fig.~\ref{fig:fittings4}.  In fact, even a decagon ($n=10$) can imitate a disk in which the error in the calculated electrical quantities is less than $<0.1\ \%$.

\subsection{Concentration of inclusions and dielectric relaxation}

By scanning the dielectric properties of mixtures in a frequency window, on gets a better understanding of the relaxation and the importance of the local electric field or in other words the interaction between inclusions. Therefore, analysis and comparison of mixtures with intermediate ($q=0.4$) and high ($q=0.7$) concentrations of inclusions were discussed in Ref.~\cite{Tuncer2001a}. The inclusions were hard disks which did not overlap; this resulted a packing density (or limiting concentration) for the model, which was $\pi/4$ for the two dimensional square lattice. The obtained frequency dependent complex dielectric permittivities were compared to analytical mixture formulas and were also curve-fitted~\cite{DataPlot} with the addition of two contributions; the Cole-Cole empirical expression~\cite{CC} and the dc conductivity,
\begin{equation}
  \label{eq:colecole}
  \eps=\epsilon+\frac{\chi(0)}{1+(\imath\omega\tau)^\alpha}+\frac{(\sigma/\eps_0)}{\imath\omega}
\end{equation}
The main reason for the curve-fitting procedure was to check the behavior of dielectric relaxation and to find a plausible explanation for deviations from exponential dielectric relaxation. The fitting procedure is presented in  Ref.~\cite{Tuncer2001a}.

\subsubsection{Intermediate concentrations of inclusions}

\begin{figure}[t]
   \centering{\includegraphics[height=3.3in,angle=-90]{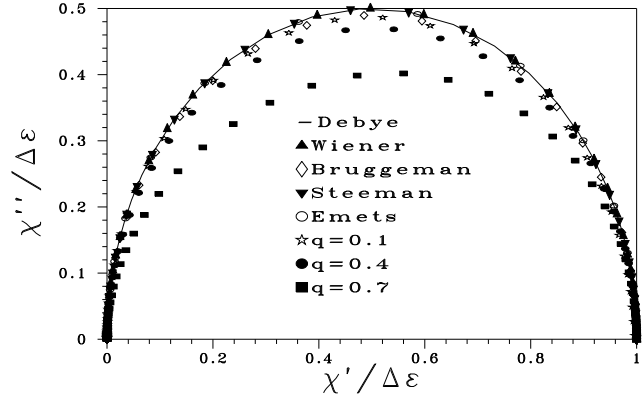}}
    \caption{\label{fig:comparison0.4}Comparison of the calculated normalized dielectric responses of a binary composites for $q=[0.1(\star),0.4(\bullet),0.7(\blacksquare)]$ and the analytical solutions for $q=0.4$. Properties of the media are $\epsilon_1=2$, $\epsilon_2=10$, $\sigma_1=10^{-12}\ \reciprocal\ohm\reciprocal\metre$ and $\sigma_2=10^{-10}\ \reciprocal\ohm\reciprocal\metre$.  The solid line ($\full$) represents a Debye-type dielectric response.}
\end{figure}

When the concentration of the inclusions phase was 0.4, the differences between different mixture formulas and the {\sc fem} solution were significant, as displayed in Fig.~\ref{fig:comparison0.4} and the curve fitting parameters are presented in Table~\ref{tab:table2}. In the figure, the dc contributions have been substructed from the data.  The analytical solution of Emets~\cite{Emets98b} (Eqs.~(\ref{eq:emets}) and (\ref{eq:delta})) was separated from the other formulas. There were no differences between the responses obtained from the Wiener~\cite{wiener} and Steeman~\cite{Steeman1992} expressions. Except for the  Bruggeman~\cite{Bruggeman1935} formula, the other dielectric responses had similar high frequency values, $\epsilon$, however, the Emets~\cite{Emets98b} solution yielded the highest value of $\eps'(\omega@1\mega\hertz)$ compared to the others. The {\sc fem} solution was close to those Wiener~\cite{wiener} and  Steeman~\cite{Steeman1992} but did not have exactly the same shape, compare the fitting parameters in Table~\ref{tab:table2}. 

The relaxation times, $\tau$, from the curve fitting procedure were close to each other except for the  Bruggeman~\cite{Bruggeman1935} response. The conductivity terms, $\sigma/\eps_0$ obtained from the curve-fitting procedure also showed that the Bruggeman~\cite{Bruggeman1935} and Emets~\cite{Emets98b} formulas yielded higher values than the others.

\begin{table}[t]
  \caption{Cole-Cole fitting parameters of different dielectric mixture formulas and the finite element calculations. The electrical parameters of the media are  $\epsilon_1=2$, $\epsilon_2=10$, $\sigma_1=1\  \pico\siemens\reciprocal\metre$ and $\sigma_2=100\ \pico\siemens\reciprocal\metre$; $q=0.4$.}
  \begin{flushright}
    \begin{tabular}{lrrrrr}
      \br
      Model  & $\epsilon$ & $\chi(0)$ & $\tau$ & $\alpha$ &$\sigma/\eps_0$\\
      \mr
      Steeman~\cite{Steeman1992} & 3.4545 & 1.0447 & 1.2689 & 1.0003 & 0.2586\\
      Bruggeman~\cite{Bruggeman1935} & 3.5785 & 1.6213 & 1.6331 & 0.9988 & 0.3031\\
      Emets~\cite{Emets98b} & 3.4881 & 1.1589 & 1.3305 & 0.9993  & 0.2677\\
      Wiener~\cite{wiener} & 3.4545 & 1.0447 & 1.2687 & 1.0003  &0.2586\\
      {\sc fem} & 3.4547 & 1.0584 & 1.2658 & 0.9897 &0.2594\\
      \br
    \end{tabular}
  \end{flushright}
  \label{tab:table2}
\end{table}

\subsubsection{High concentration of inclusions}

When the concentration of the inclusions was  $0.7$ which is close to the limiting concentration of the square lattice, $q=\pi/4=0.7854$, only the Wiener~\cite{wiener} and  Steeman~\cite{Steeman1992} responses were similar to each other, as presented in Table~\ref{tab:table3}. All the responses were different, and the high frequency dielectric permittivities $\epsilon$ were between $5.5$ and $6$. The low frequency values of permittivity, $\varepsilon'(\omega@1\micro\hertz)$, on the other hand, had large discrepancies. It was extraordinary that the Emets~\cite{Emets98b} formula resulted in the highest value of the real part of the dielectric permittivity at low frequencies, $\epsilon+\chi(0)$, while in the other considered cases the response estimated by Bruggeman~\cite{Bruggeman1935} had this character. 
The {\sc fem} solution was non-symmetric, and it could not be modeled with the Cole-Cole empirical formula as presented in Fig.~\ref{fig:comparison0.4}.   The fitting parameters of the {\sc fem} have shown that the obtained reponse did not have any Cole-Cole characteristics. 

Thus, for a simple computational domain of a binary structure, the dielectric relaxation was Debye-like when the concentration of the inclusions was low, and  the mixture formulas of Steeman~\cite{Steeman1992}, Wiener~\cite{wiener} and Emets~\cite{Emets98b} agreed with the {\sc fem} solutions~\cite{Tuncer2001a}. Non-Debye and non-Cole-Cole dielectric relaxations were observed for the {\sc fem} solutions at higher inclusion concentrations. Since a simple binary system composed of a square lattice exhibited non-expected dielectric responses, introducing geometries and structures which remind more realistic situations would  probably yield non-Debye like responses as well. When the dimensions and shapes of the inclusions were different, the time constant of the relaxation would not be single but distributed.
  \begin{table}[t]
    \caption{Cole-Cole fitting parameters of different dielectric mixture formulas and the finite element calculations. The electrical parameters of the media are  $\epsilon_1=2$, $\epsilon_2=10$, $\sigma_1=10^{-12}\ \reciprocal\ohm\reciprocal\metre$ and $\sigma_2=10^{-10}\ \reciprocal\ohm\reciprocal\metre$; $q=0.7$.}
    \begin{flushright}
      \begin{tabular}{lrrrrr}
\br
Model  & $\epsilon$ & $\chi(0)$ & $\tau$ & $\alpha$ &$\sigma/\eps_0$\\
\mr
Steeman~\cite{Steeman1992} & 5.4998 & 4.7154 & 1.7875 & 1.0003 & 0.6068\\
Bruggeman~\cite{Bruggeman1935} & 5.8779 & 10.056 & 3.4816 & 0.9651 & 1.0528\\
Emets~\cite{Emets98b} & 6.0313 & 10.624 & 3.1293 & 0.9975  & 1.0697\\
Wiener~\cite{wiener} & 5.4998 & 4.7153 & 1.7875 & 1.0003  &0.6068\\
{\sc fem} & 5.7991 & 6.9035 & 2.1502 & 0.9609 & 0.7685\\
\br
      \end{tabular}
    \end{flushright}
    \label{tab:table3}
\end{table}

\subsection{Structural differences and dielectric relaxation}
\subsubsection{Ordered structures with different lattices}

\begin{figure}[t]
  \begin{center}
      {\includegraphics[angle=-90,width=1.5in]{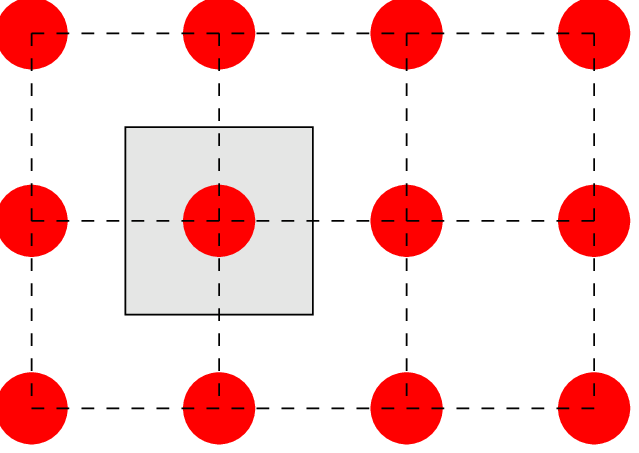}}
      {\includegraphics[angle=-90,width=1.5in]{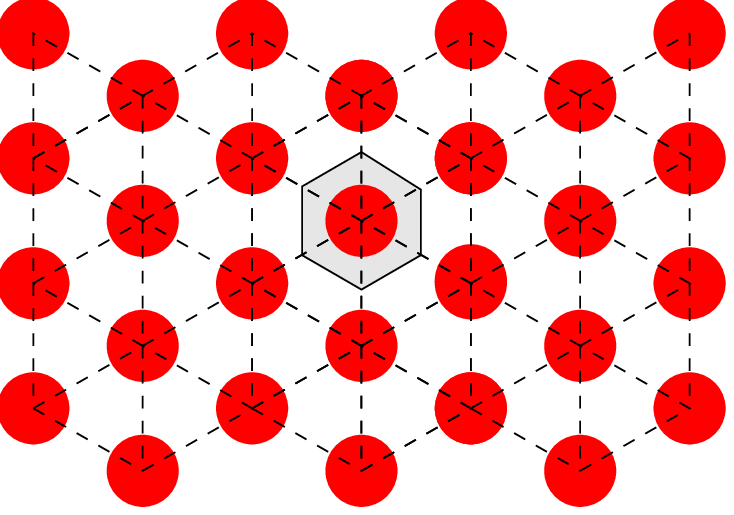}}
  \end{center}
  \caption{\label{fig:geom} Square (left) and   triangular (right) lattices with square and hexagonal unit cells. Dark circles are the inclusions which are distributed in the host medium (white background). The shaded areas are the primitive (unit) cells which are used in the calculations.}    
\end{figure}

The considered regular lattices were square and triangular as presented in Fig.~\ref{fig:geom}. In these structures, the smallest repeating units used in the calculations are shown as shaded areas in the figure. Two different cases, in which the inclusions were more ($\sigma_1<\sigma_2$ -- match-composite) and less ($\sigma_1>\sigma_2$ -- reciprocal-composite) conductive than the matrix phase were taken into consideration. The ohmic conductivities of the media were either $1\ \pico\siemens\per\meter$ or $100\ \pico\siemens\per\meter$.

\begin{figure}[t]
  \begin{center}
    {\includegraphics[height=3.3in,angle=-90]{REGMC.EPS}}
    {\includegraphics[height=3.3in,angle=-90]{REGFMC.EPS}}
  \end{center}
    \caption{\label{fig:regsqdistmc} (Up) Product of dielectric strength, $\Delta\eps$, and distribution of relaxation times, $\mathcal{F}$, found using the Monte Carlo technique~\cite{Tuncer2000b}. The open ($\circ-\Box$) and filled ($\bullet-\blacksquare$) symbols represent the responses when $\sigma_1<\sigma_2$ (match-composite) and $\sigma_1>\sigma_2$ (reciprocal-composite), respectively. The boxes ($\boxdot-\blacksquare$) and circles ($\circ-\Box$) represent the square and triangular lattice, respectively. (Down) Reconstructed dielectric responses, the chain lines ($\chain$), using the distributions in upper subfigure.}
\end{figure}

The analysis of the dielectric responses by means of the Monte Carlo technique~\cite{Tuncer2000b} are presented in Fig.~\ref{fig:regsqdistmc}a. The product of dielectric strength and the distribution of relaxation times, $\Delta\eps\times\mathcal{F}$, of the $\sigma_1<\sigma_2$ case, ($\odot-\boxdot$), for two different regular structures were similar and nearly-symmetrical. The slopes of the distributions could be fitted by $(\log\tau)^\kappa$ expression, where $\kappa\approx6$ for $\log\tau<0$, and $\kappa\approx-5$ for $\log\tau>0.3$. It is possible to convolute this distribution into two separate ones characterized by different dominant time constants. When $\sigma_1>\sigma_2$, ($\bullet-\blacksquare$), the distributions were broader. Moreover, the slope of the distributions at shorter $\tau$-values were identical but, the arrangements of the inclusions had altered the distribution of relaxation times at longer times, $\log\tau>-0.1$. The reconstructed dielectric responses are presented in Fig.~\ref{fig:regsqdistmc}b and they were more successful than the Cole-Cole curve-fittings.

\subsubsection{Disordered structures}
\begin{figure}[t]\centering{
    \begin{tabular}{cc}
      {\includegraphics[width=1.5in]{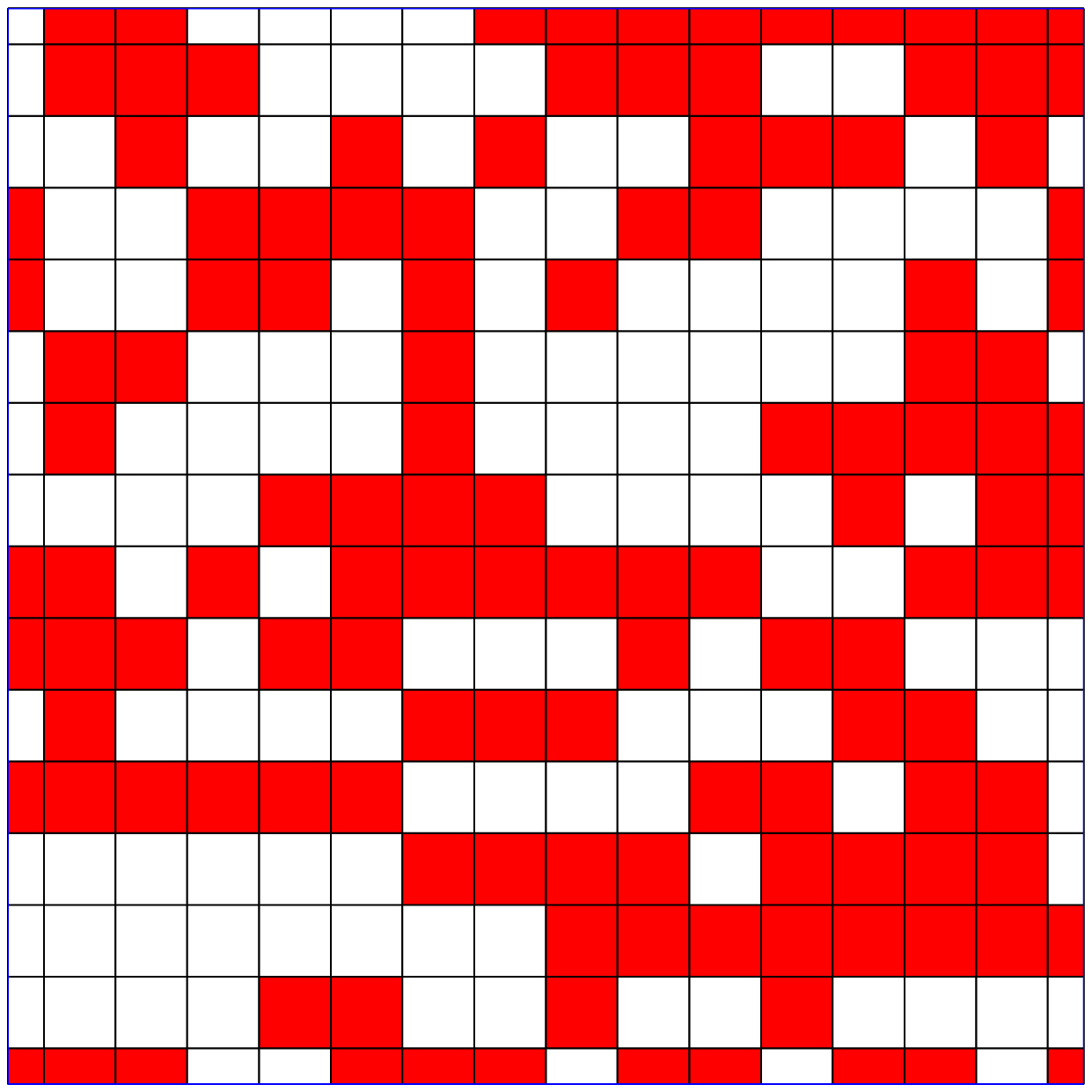}}&
      {\includegraphics[width=1.5in]{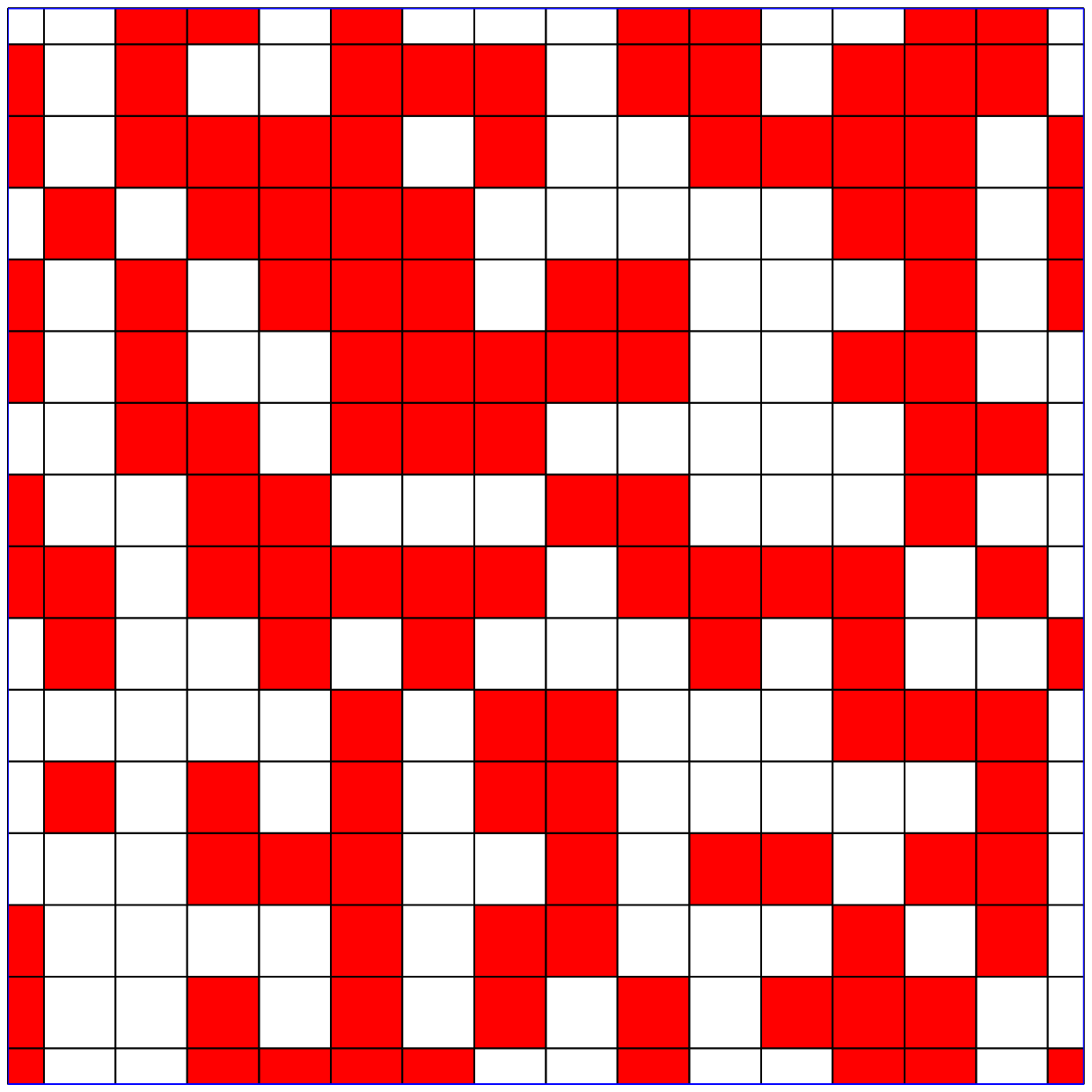}}\\
      (a) &(b) \\ 
      {\includegraphics[width=1.5in]{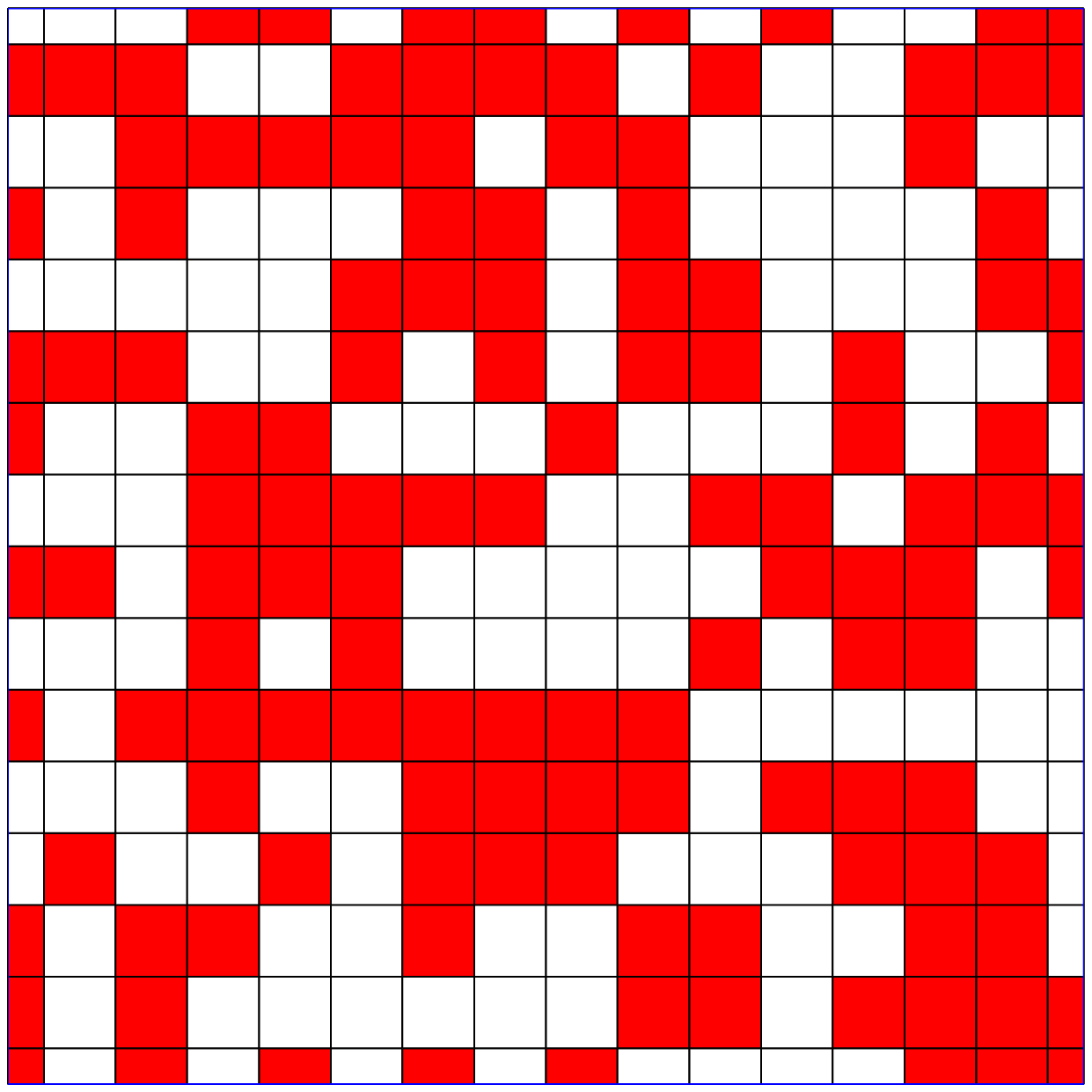}}&
      {\includegraphics[width=1.5in]{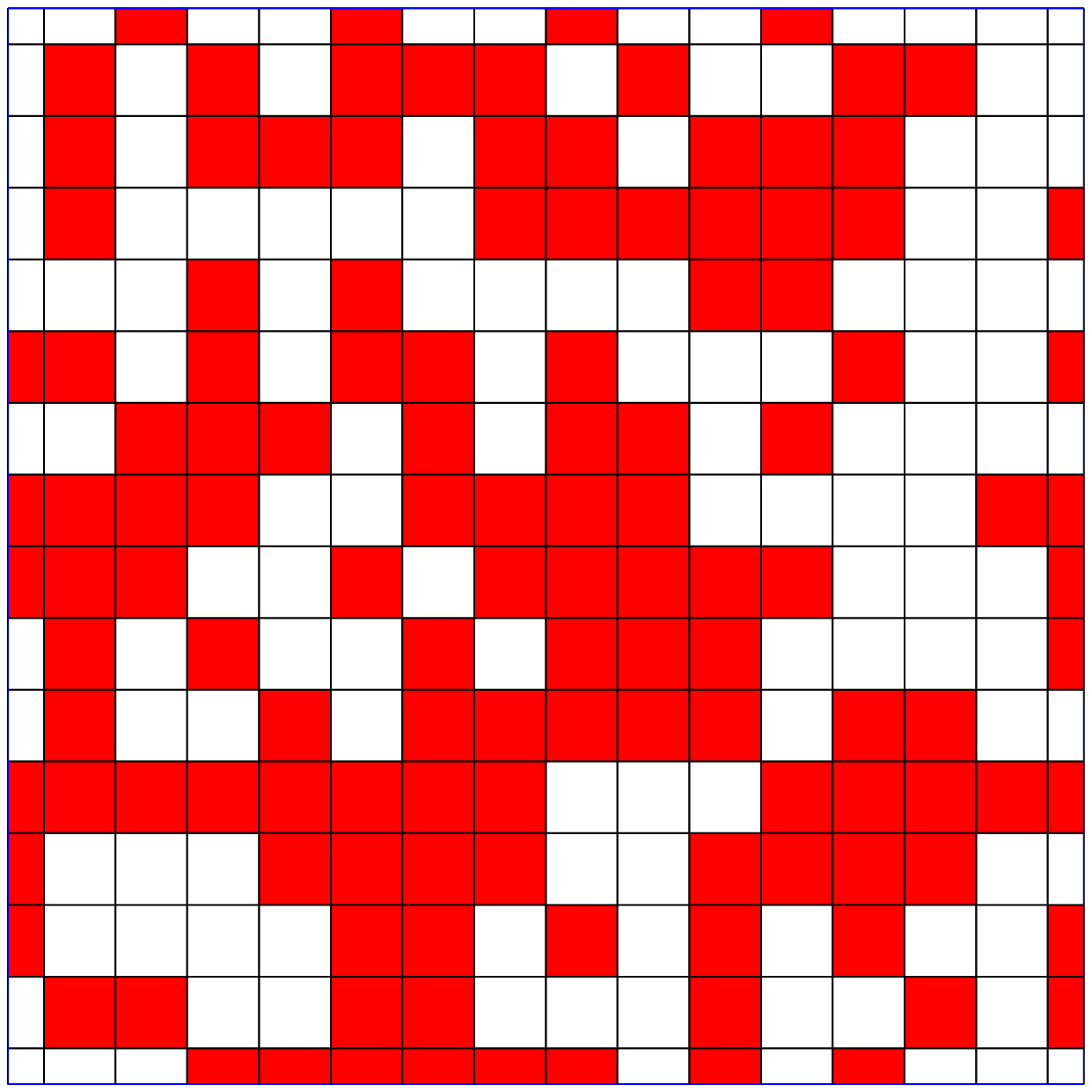}}\\
      (c) &(d) 
    \end{tabular}}
  \caption{Structures in which the minimum, $\eps'(w@200\pi\ \hertz)$, (a\&c) and the maximum, $\eps'(w@2\pi\ \milli\hertz)$, (b\&d) dielectric permittivities are obtained when $\sigma_1<\sigma_2$ (a\&b) and $\sigma_1>\sigma_2$ (c\&d).\label{fig:rand_sq}\label{fig:geom2dis}}
\end{figure}

Earlier numerical simulations~\cite{SihvolaBook,Tuncer1998b,TuncerPhD,Tuncer2001a,Sareni1996,shen1990,sar97,BoJMS} of dielectric mixtures by means of the {\sc fem}~\cite{Tuncer1998b,TuncerPhD,Tuncer2001a,Sareni1996,sar97} have not considered frequency dependent properties of disordered systems. However, some experimental and theoretical investigations have been reported~\cite{Clerc1990,Lux1993,Basu,Dissado1,Vainas}. The authors of this report, as in regular lattice structures investigated behavior of random systems with concentrations of the phases constant, $q_{1,2}=0.5$. A concept by of puzzle-like structures was introduced, as presented in Fig.~\ref{fig:geom2dis}. The networks were composed of $16\times16$ cells, and the matrix and inclusion elements were placed randomly. Altogether 1024 structures were analyzed. A range of responses were obtained depending on the ratio of cunductivity values of the structure constituents as well as their distribution.

\begin{table}[t]
  \caption{Curve fitting parameters of Eq.~(\ref{eq:havneg}) using Cole-Cole and Davidson-Cole empirical expressions for the dielectric relaxation of random structures with square networks.\label{tab:table_rndsq}}
  \begin{flushright}
    \begin{tabular}{lcccccc}
      \br
      Structure & case &$\epsinfa$&$\Delta\eps$&$\tau[\second]$&$\alpha$&$\sigma/\eps_0[\siemens\per\farad]$\\
      \mr
      Fig.~\ref{fig:rand_sq}a & $\sigma_1<\sigma_2$& 4.145 &4.819 &3.916 &0.921&0.609\\
      Fig.~\ref{fig:rand_sq}b & $\sigma_1<\sigma_2$& 4.654 &9.260 &6.303 &0.927&1.185\\
      Fig.~\ref{fig:rand_sq}c & $\sigma_1>\sigma_2$& 4.424 &33.89 &9.329 &0.751&1.623\\
      Fig.~\ref{fig:rand_sq}d & $\sigma_1>\sigma_2$& 4.803 &65.29 &23.49 &0.818&1.159\\
      \mr
      &  & & & &$\beta$ & \\
      \mr
      Fig.~\ref{fig:rand_sq}a & $\sigma_1<\sigma_2$& 4.134 &4.796 &5.045 &0.818&0.609\\
      Fig.~\ref{fig:rand_sq}b & $\sigma_1<\sigma_2$& 4.630 &9.222 &8.068 &0.822&1.185\\
      Fig.~\ref{fig:rand_sq}c & $\sigma_1>\sigma_2$& 4.085 &33.44 &23.88 &0.523&1.623\\
      Fig.~\ref{fig:rand_sq}d & $\sigma_1>\sigma_2$& 4.213 &64.66 &46.62 &0.600&1.159\\
      \br
    \end{tabular}
  \end{flushright}
\end{table}

\begin{figure}[b]
  \centering{\includegraphics[height=3.3in,angle=-90]{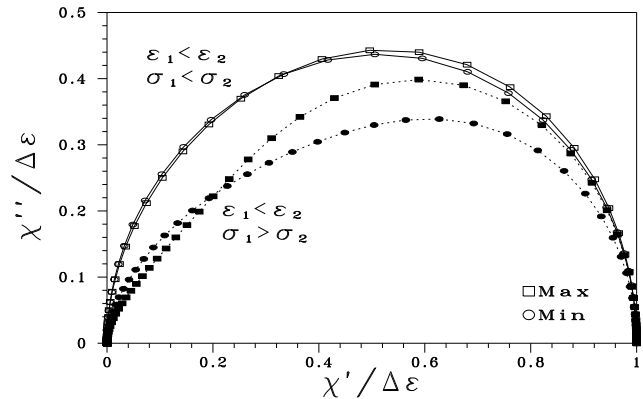}}
  \caption{\label{fig:rand_relax_sq}Normalized dielectric responses of random structures with square network. The minimum and maximum dielectric permittivities obtained when $\sigma_1<\sigma_2$ (Fig.~\ref{fig:rand_sq}a \& b) and $\sigma_1>\sigma_2$ (Fig.~\ref{fig:rand_sq}c \& d). The open ($\circ-\Box$) and filled ($\bullet-\blacksquare$) symbols represent the responses of structures in Fig.~\ref{fig:rand_sq} in which the lowest, $\varepsilon(w@200\pi\ \hertz)$, and the highest, $\varepsilon(w@2\pi\ \milli\hertz)$, dielectric permittivities are obtained, respectively.}
\end{figure}

Examples of the frequency dependent dielectric permittivities of the structures are displayed normalized in  Fig.~\ref{fig:rand_relax_sq} after subtracting of the ohmic conductivity  contributions.  The effect of randomness was significant compared to the regular structures, the influence of the local field enhancement and internal structure yielded stronger polarizations.  The curve fitting parameters given in Table~\ref{tab:table_rndsq} have indicated that there were slight differences between the two empirical formulas for the $\sigma_1<\sigma_2$ case (Fig.~\ref{fig:rand_relax_sq}a), however for the $\sigma_1>\sigma_2$ case (Fig.~\ref{fig:rand_relax_sq}b) the Davidson-Cole expression produced a more successful fit (lower residual) than the Cole-Cole one.

All the obtained responses started and ended at the $\eps'$-axis with an angle of $\pi/2$, which indicated that they were neither Cole-Cole nor Davidson-Cole; a Cole-Cole response starts with $\alpha\pi/2$ and, similarly, a Davidson-Cole response starts with $\beta\pi/2$ in the Cole-Cole representation. This has shown that there exist two relaxation time constants, $\tau$,  which set boundaries for the distribution of relaxation times~\cite{MacDonald1987}. The reconstructed response using the distribution of relaxation times is displayed by lines in Fig.~\ref{fig:rand_relax_sq}.

The analysis has showed that the randomness in the medium spread the dielectric relaxation times over a broder region compared to the regular structures. If the finite-size effects had been taken into account, assuming that the calculations were done on larger crosswords (not $16\times16$), or if our results could have been extrapolated, the dielectric relaxation in random media would yield a behavior reminding the low frequency dispersion~\cite{Jonscher1983,HillLFD,jons92}.

\subsubsection{Microstructural differences in random lattices}
\begin{figure}[tp]
  \begin{center}
      \psfragscanon
      \psfrag{FRAME16}{}
      {\includegraphics[height=1.5in]{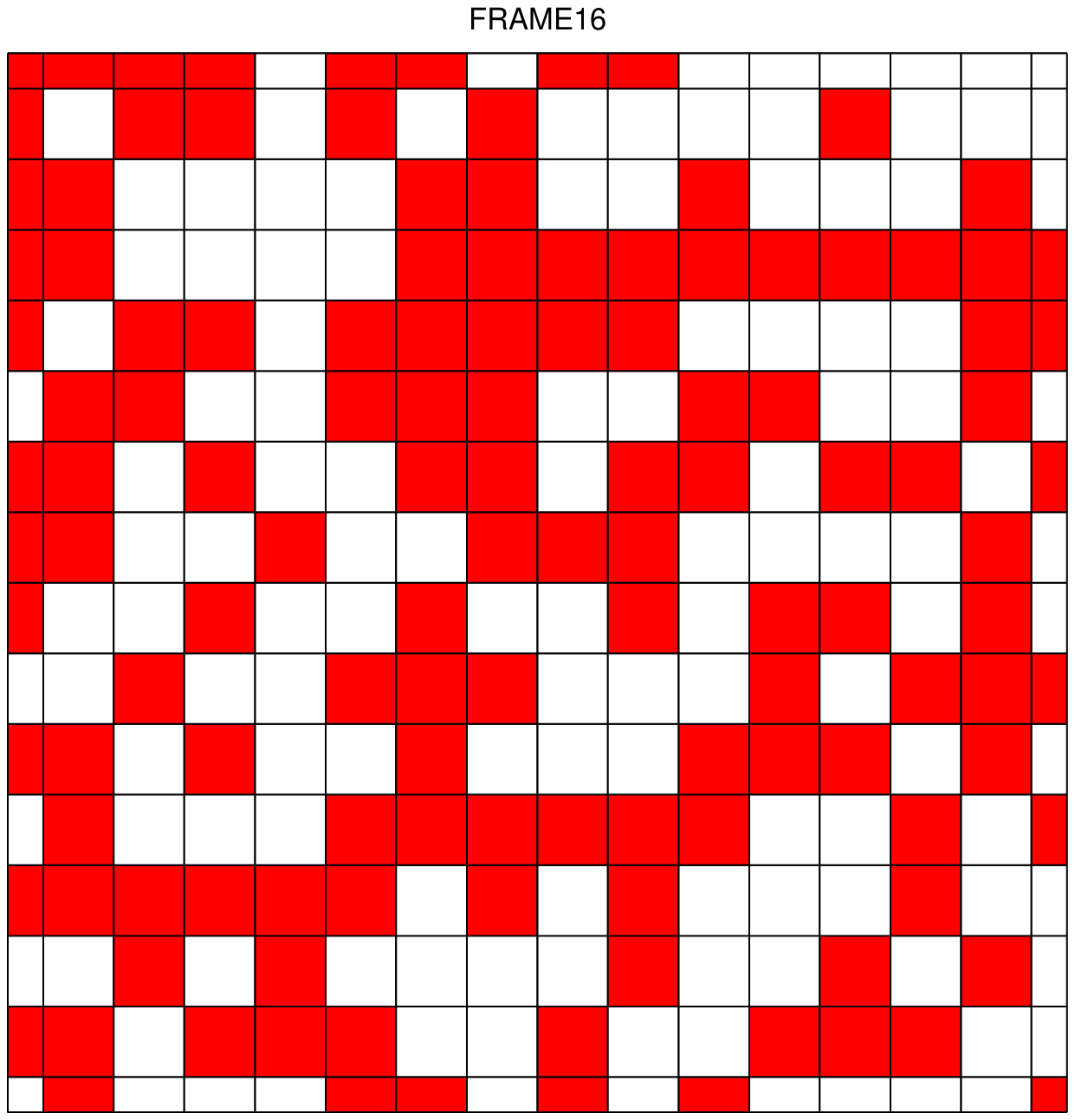}}
      {\includegraphics[height=1.5in,width=1.7in]{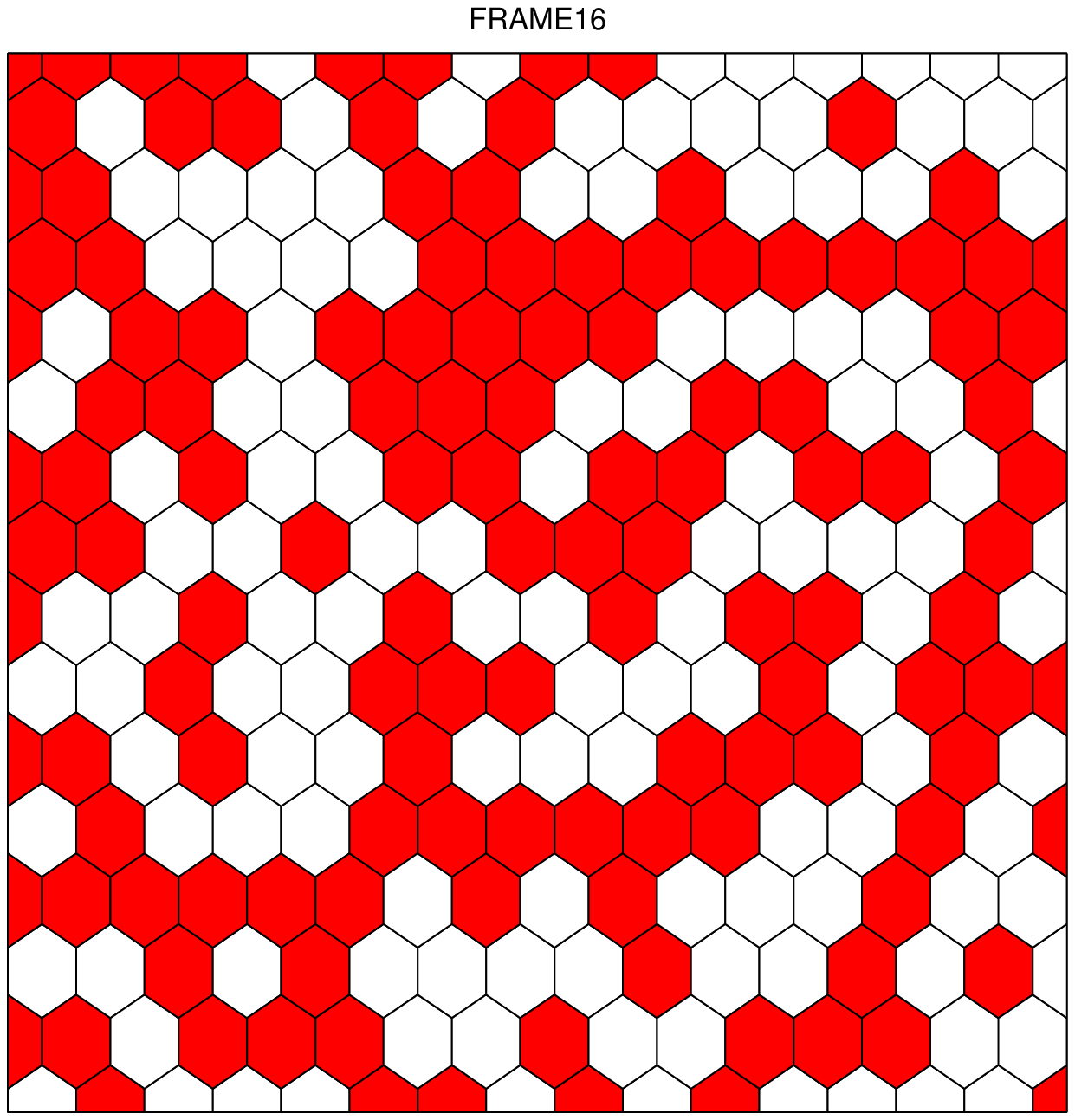}}
      \psfragscanoff
  \end{center}
  \caption{\label{fig:geom2}Square (left) and  triangular (right) $16 \times 16$ lattice structures corresponding to the same set of tile topology. The dark and white polygons are phases with different electrical properties.}
\end{figure} 

\begin{figure}[t]
  \centering{\includegraphics[height=3.3in,angle=-90]{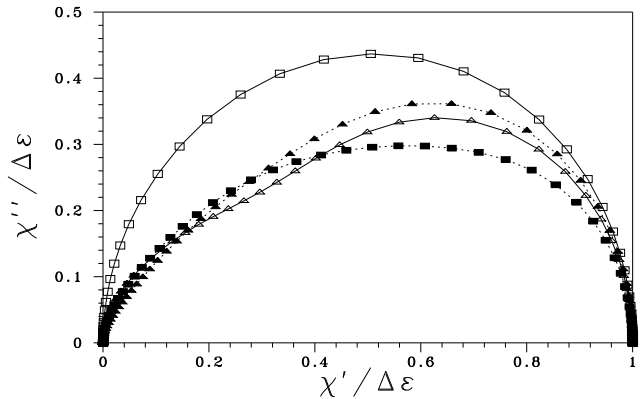}}
  \caption{ \label{fig:res6}Normalized dielectric responses obtained for two chosen random distributions with square ($\Box-\blacksquare$) and honeycomb ($\vartriangle-\blacktriangle$) tiles.  The open ($\Box-\vartriangle$) and filled ($\blacksquare-\blacktriangle$) symbols represent the real, $\chi'$ and imaginary, $\chi''$ parts of the calculated dielectric susceptibility.}
\end{figure}

When random structures with different lattices, as shown in Fig.~\ref{fig:geom2}, were considered, square and honeycomb tiles the smallest parts. The frequency dependent complex dielectric susceptibilities of selected structures with different lattices having the same distribution of phases are presented in Fig.~\ref{fig:res6}.  The shape of the Cole-Cole plots of the obtained responses after the substructing of the ohmic losses were non-exponential, similar to those presented previously in Fig.~\ref{fig:rand_relax_sq}, as presented in Fig.~\ref{fig:res6}. In the figure, the reponses were for the reciprocal-composites ($\sigma_1>\sigma_2$), and in the investigation two different conductivity ratios of the constituents were taken into account. Only the structure with square tiles had symmetrical response for $\varepsilon_1/\varepsilon_2=1/5$ and $\sigma_1/\sigma_2=100$, the other reponses were asymmetrical. When the ratio of the conductivities were increased, $\sigma_1/\sigma_2=1000$, the dielectric relaxation of structure with square tiles became distorted. It was interesting that the structure with honeycomb tiles had similar dielectric responses for two considered cases with different conductivity ratios. When the conductivity ratio was increases, a condition for the `normal' low frequency dispersion~\cite{Jonscher1983,Dissado1,HillLFD} was created since both responses had flat regions. The ohmic conductivities of the structures were different and the structure with honeycomb tiles had higher conductivity values.  

\section{Discussion and future challenges}

In dilute systems, the dielectric relaxation due to interfacial polarization yields classical Debye relaxation, which was also verified by numerical simulations. However for dense mixtures non-Debye (non-exponential) relaxation behavior can be observed depending on the internal structure of the system and the electrical properties of the constrituents. The differences between regular and random structure were remarkable, such that, the random structures had broader non-Debye dielectric relaxations when the complex dielectric permittivity ratio of the phases were small.  Further modification of the constituents' properties indicated that, it was even possible to obtain a dielectric response remanding the low frequency dispersion behavior. Therefore, one should be careful in interpretating dielectric data since the interfacial polarization could result in various dielectric relaxations depending on the internal structure of the mixture and on the properties of the constituents.

The most important question arising which concerns any kind of numerical simulations is the reliability of obtained results. Here the reliability does not mean the correctness of a physical model representing a real physical phenomenon, but its mathematical and computer implementations. In the case of modeling dielectric properties of composites, this is one of the basic problems because usually there are no exact analytical solutions avaliable that would allow for  checking the accuracy of the computed data. Special attention should be paid, for example, when a method for calculating the effective complex permitivity is chosen. Four different methods mentioned in \S~\ref{sec:appl-finite-elem} were compared, but only two of them (one based on the total current and phase shift between the current and applied voltage; and one where average field and dielectric displacement were calculated) yielded stable and reliable solutions~\cite{Tuncer-CEIDP01,TuncerAcc1}. Physical (or mathematical) reasons of discrepancies of the data obtained with different methods are not completely understood yet. 

It is obvious, that correct numerical representation of the initial problem must be used in simulations and appropriate meshing of the computational domain must be performed. Meshing is extremely important for the considered problem in order to resolve anisotropy in the composite structure. In this connection, special care should be taken about meshing ability of the used software for solving the field problem in Eq.~(\ref{eq:contin.2}), because it could be very difficult and even impossible to generate a computational mesh if great differences in the dimensions of the matrix and inclusions exist. 

The desired outcome from simulations is an agreement between computed results and characteristics of a real composite material. That would prove the applicability of the model and allows performing simulations for other structures with unknown effective parameters. This can be achieved if a real three-dimensional structure of the material is built, and the dependences of electrical properties of the constituents on state variables (frequency, temperature, pressure, electric field, {\em etc.}) are introduced in simulations. Furthermore, a possibility of modeling different properties of interfacial layers between constituents of the composite should be introduced. These problems have not been solved yet, and is the subject of on-going research.

\section*{References}

\bibliography{../../../references}
\bibliographystyle{is-unsrt}
\end{document}